\documentclass[journal]{IEEEtran}
\renewcommand\IEEEkeywordsname{Keywords}
\usepackage{amssymb}
\usepackage{graphicx}
\usepackage{stfloats}
\usepackage{placeins}
\usepackage{subfig}
\usepackage{longtable}
\usepackage{floatrow}
\usepackage{float}
\usepackage{multirow}
\usepackage{amsmath}
\usepackage{amssymb}
\usepackage{amsfonts}       
\usepackage{fixltx2e}
\usepackage{tabularx}
\usepackage{paralist,array}
\usepackage{paralist}
\usepackage{array}
\usepackage{textcomp}
\usepackage{float}
\usepackage{multirow}
\usepackage{algorithm,algorithmic}       
\usepackage[T1]{fontenc}
\usepackage{cite}
\graphicspath{{Figures/}}
\usepackage[autostyle]{csquotes}
\usepackage{tikz}
\usetikzlibrary{shapes,shapes.geometric,arrows,fit,calc,positioning,automata}
\usepackage[autostyle]{csquotes}
\newcounter{saveenumi}

\makeatletter
\newcommand{\ALOOP}[1]{\ALC@it\algorithmicloop\ #1%
  \begin{ALC@loop}}
\newcommand{\ENDALOOP}{\end{ALC@loop}\ALC@it\algorithmicendloop}

\makeatother

\begin{document}
%
\title{Joint Bilateral Filter for Signal Recovery from Phase Preserved Curvelet Coefficients for Image Denoising}

\author{Supratim~Gupta,~and~Susant~Kumar~Panigrahi
        \thanks{S. Gupta (e-mail: sgupta.iitkgp@gmail.com) and S. K. Panigrahi (e-mail: susant146@gmail.com) are with the Department of Electrical Engineering, National Institute of Technology, Rourkela, Odisha-769008, India.}}


\maketitle

\begin{abstract}
Thresholding of Curvelet Coefficients, for image denoising, drains out subtle signal component in noise subspace. This produces ringing artifacts near edges and granular effect in the denoised image. We found the noise sensitivity of Curvelet phases -- in contrast to their magnitude -- reduces with higher noise level. Thus, we preserved the phase of the coefficients below threshold at coarser scale and estimated their magnitude by Joint Bilateral Filtering (JBF) technique from the thresholded and noisy coefficients. In the finest scale, we apply Bilateral Filter (BF) to keep edge information. Further, the Guided Image Filter (GIF) is applied on the reconstructed image to localize the edges and to preserve the small image details and textures. The lower noise sensitivity of Curvelet phase at higher noise strength accelerate the performance of proposed method over several state-of-the-art techniques and provides comparable outcome at lower noise levels.
\end{abstract}


\IEEEpeerreviewmaketitle

\section{Introduction}
\label{Sec1:Introduction}
Digital image is corrupted during acquisition, transmission and reception systems by different noises. The resultant noise due various sources in these situations can be modeled as additive Gaussian noise (AWGN). Thus the noisy image $y$ may be represented as follows:

\begin{equation}
\label{Eq:Sec1:Noise_Model}
y(\mathcal{P}) = z(\mathcal{P}) + \eta(\mathcal{P})
\end{equation}

Where, the latent image, $z(\mathcal{P})$ is contaminated with uncorrelated, zero mean additive white Gaussian noise (AWGN), $\eta \in \aleph({0, {\sigma}^2})$ to produce noisy image $y(\mathcal{P})$ on a $2D$ grid of $\mathcal{P} \in {\mathbb{R}}^2$. Out of several attempts  to recover the latent image, `Sparse-Land' signal modeling and non-local (and/or local) patch based image synthesis in spatial domain are proved to be most powerful tools.

In `Sparse-Land' modeling, an image is represented with minimum number of non-zero coefficients (sparse) in multiple scales. Thus a well estimated threshold -- at different scales, $\gamma$ -- may be able to separate signal from noise subspace by keeping coefficients of higher magnitude. The thresholded coefficients ($\hat{Y}_{\gamma}$) is transformed back into spatial domain to produce the denoised image, $\hat{z}(\mathcal{P})$ as formulated in Eq.\ref{Eq:Sec1:ThreshDenoising}.

\begin{subequations}\label{Eq:Sec1:ThreshDenoising}
     \begin{align}
     Y_{\gamma} &= \mathcal{\textbf{T}}_{\gamma}\left[y(\mathcal{P})\right]\label{Eq:Sec1:NoiseImage_ForwardTransform}\\
     \hat{Y}_{\gamma} &=\begin{cases}
    Y_{\gamma}~, & \text{if\, $|Y_{\gamma}| > {\lambda_\gamma}$}\\
    0~, & \text{otherwise}
        \end{cases}\label{Eq:Sec1:Hard_Threshold}\\
\hat{z}(\mathcal{P}) &= \mathcal{\textbf{T}}_{\gamma}^{-1}\left[\hat{Y_{\gamma}}\right]\label{Eq:Sec1:DenoisedIm_InverseTransform}
     \end{align}
\end{subequations}

Among a number of image transformation techniques, researchers had used redundant or non-redundant wavelet dictionaries for denoising and kept the salient image features like edges \cite{donoho1994ideal,donoho1995wavelet,chang2000adaptive,sendur2002bivariate,pizurica2006estimating,dengwen2008image,luisier2010sure,portilla2003image,lazzaro2007edge,tekin2013benefits}.
Recently, Curvelet transform is observed to represent images with edges, even more sparsely \cite{starck2002curvelet,ma2007combined,ma2010curvelet,remenyi2014image}. In this technique, an image is decomposed into magnitude and phase at multiple scales and directions. A hard threshold based approach was reported in \cite{starck2002curvelet} to separate the signal from the noise subspace. Here, the threshold is computed from and applied to the magnitude of the Curvelet coefficients to recover the signal from the noisy observation. This removes the corresponding phases too. Moreover, portions of the signal components may spilled into noise subspace and are lost. The hard thresholding also introduces ringing artifacts around the edges due to sudden jump in coefficient magnitudes. At the finest scale signal and noise magnitudes are comparable. Therefore, removal of coefficients with thresholds introduces granular artifacts in the denoised image.

\begin{figure*}[ht]
    \centering
    \subfloat[\label{MagSensitivityPlot_AWGN}]{\includegraphics[scale=0.5]{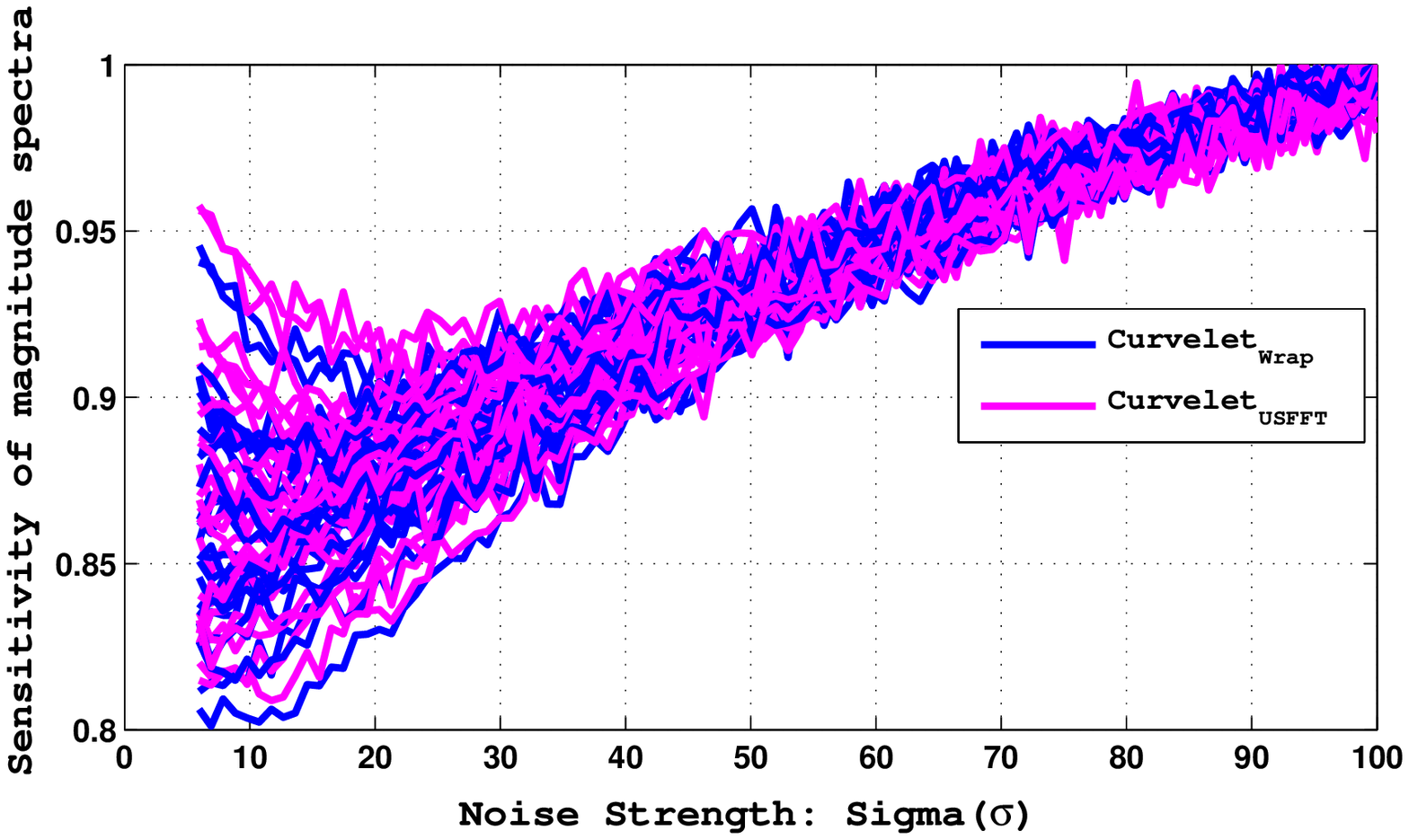}}
    \hspace{0.4cm}
    \subfloat[\label{PhaseSensitivityPlot_AWGN}]{\includegraphics[scale=0.5]{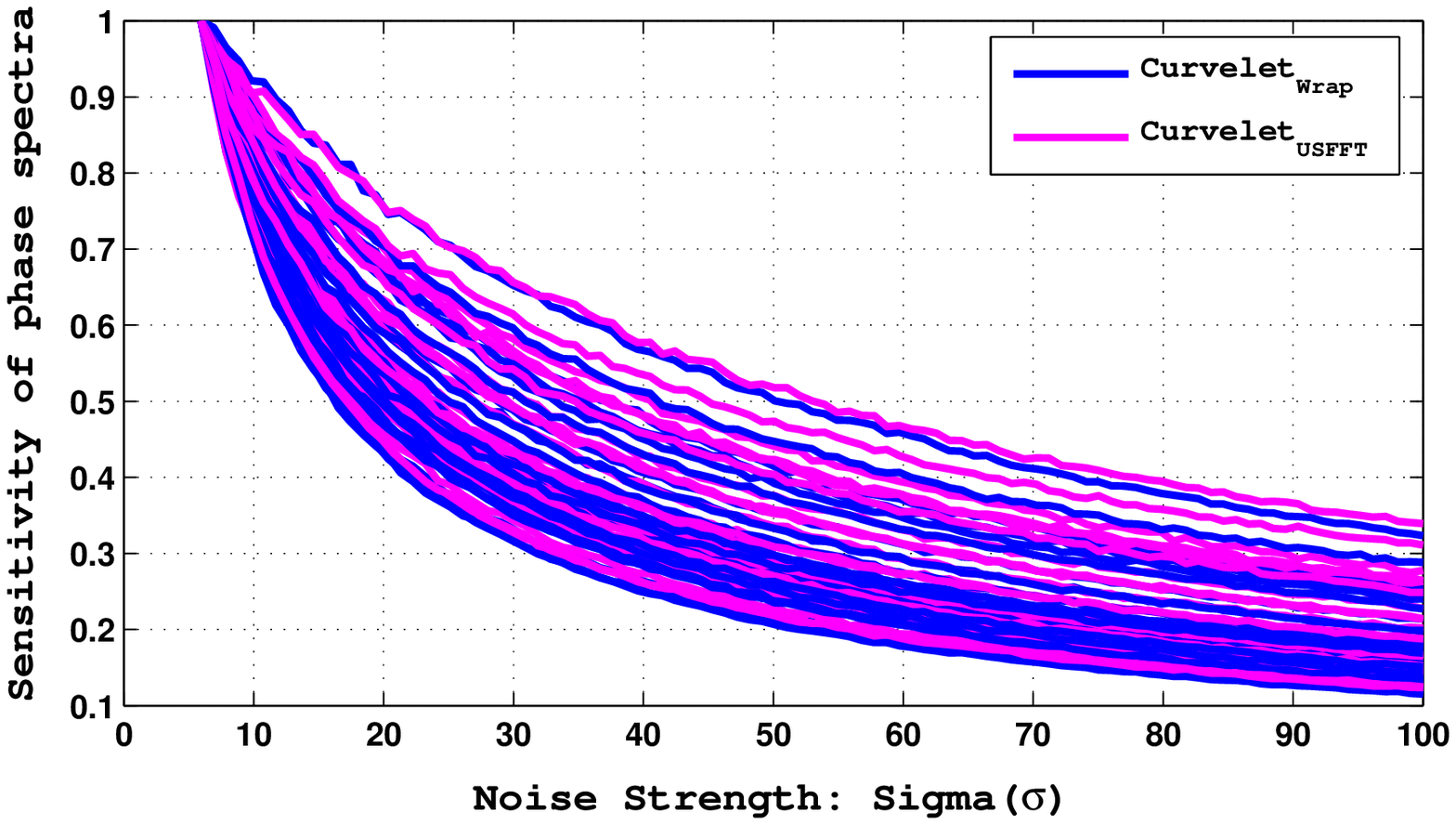}}
    \caption{\small{Noise sensitivity (Normalized between $[0, 1]$) of (a) Magnitude \& (b) Phase for Curvelet Transform -- Wrapping and USFFT -- on the images of TID$2008$ Database \cite{ponomarenko2009tid2008}.}}
    \label{Fig:Sec:Noise_SensitivityPlot}
\end{figure*}

Unlike frequency domain approaches, modern spatial filters extract self similarities among local or non-local regions for denoising \cite{tomasi1998bilateral,buades2005review,takeda2007kernel,yaroslavsky2012digital}. Among these methods, Bilateral Filter \cite{tomasi1998bilateral} and its variants \cite{milanfar2013tour} -- while retain edges effectively -- fail to restore smooth regions . This low contrast region can be recovered by determining the chromatic distance of bilateral filter from additional image different from the image of concern. This leads to Joint Bilateral Filter for image restoration \cite{petschnigg2004digital}.

Recently, several methods utilized the advantages of one domain to improve the denoising performance in another domain \cite{zhang2008multiresolution,ShreyamshaKumar2013,wu2014curvelet,zhang2016image}. Multiresolution Bilateral Filter (MBF) implements BF in the approximation scales to suppress the coarser-grain (low frequency) noise and thresholds the wavelet coefficients in detail scales to remove the fine-grain (high frequency) noise \cite{zhang2008multiresolution}. Similar to MBF, DDID \cite{knaus2013dual} integrated Joint Bilateral Filter (JBF) and Short Time Fourier Transform (STFT) based wavelet shrinkage technique for image denoising. Furthermore, Kumar \emph{et al.} \cite{kumar2013image} applied wavelet thresholding to recover the lost image detail from the residual image of NLM filter. The inability of wavelets in handling curve singularity may limit the performance of these hybrid techniques. The Non Subsampled Shearlet Transform (NSST) -- in contrast to wavelet -- represents image along multiple directions \cite{easley2008sparse}. A linear combination of local Wiener filter and its method noise thresholded image using NSST is proposed in \cite{zhang2016image} to recover the image from its noisy observation. However the selection of parameters for linear combination in \cite{zhang2016image} is intuitive, which may limit its performance for wide variety of natural images. In contrast with several variants of combined approaches, block matching 3D collaborative filtering (BM3D) technique excelled in denoising by grouping the similar (non-local) patches and collaboratively filtering the 3D blocks using 1D wavelet thresholding \cite{dabov2007image}. The state-of-the-art BM3D technique still inadequate in restoring few homogenous regions that manifests as low-frequency noise \cite{knaus2014progressive}.

We found the noise sensitivity of Curvelet phase  -- in contrast to its magnitude -- reduces with higher  noise level.  This indicates preserving the phase of Curvelet coefficients may improve denoising quality. We (re)estimated the signal magnitude in the coarser scales by implementing JBF on the thresholded coefficients. This process recovers the signal residuals from the noise subspace, while retaining the indispensable phase information. On the other hand, the implementation (fast) BF (using Fourier kernels \cite{ghosh2016fast}) in the finest scale ensures the elimination of granular artifacts with well-connected edges in the restored image. Finally, the reconstructed image is further processed using GIF ($\mathcal{O}(M)$) for better preservation of local structures like: edges, textures and small details. The performance of the proposed algorithm is measured using a few statistical and edge localization measures to compare its efficacy with several state-of-the-art techniques.

The rest of the article is organized as follows: Section~\ref{Sec2:NoiseDistribution} provides few justification on the use of JBF and BF in Curvelet domain and also illustrates the noise sensitivity of the phase and the magnitude under additive Gaussian noise (AWGN). The proposed method for image denoising is explained in Section~\ref{Sec3:ProposedMethod}. The experimental results are given and discussed in Section~\ref{Sec:Results}. Finally, Section~\ref{Sec:Conclusion} concludes the paper.

\begin{figure*}[!ht]
    \centering
    \subfloat[\label{PDF_CTsc2or1GaussN}]{\includegraphics[scale=0.51]{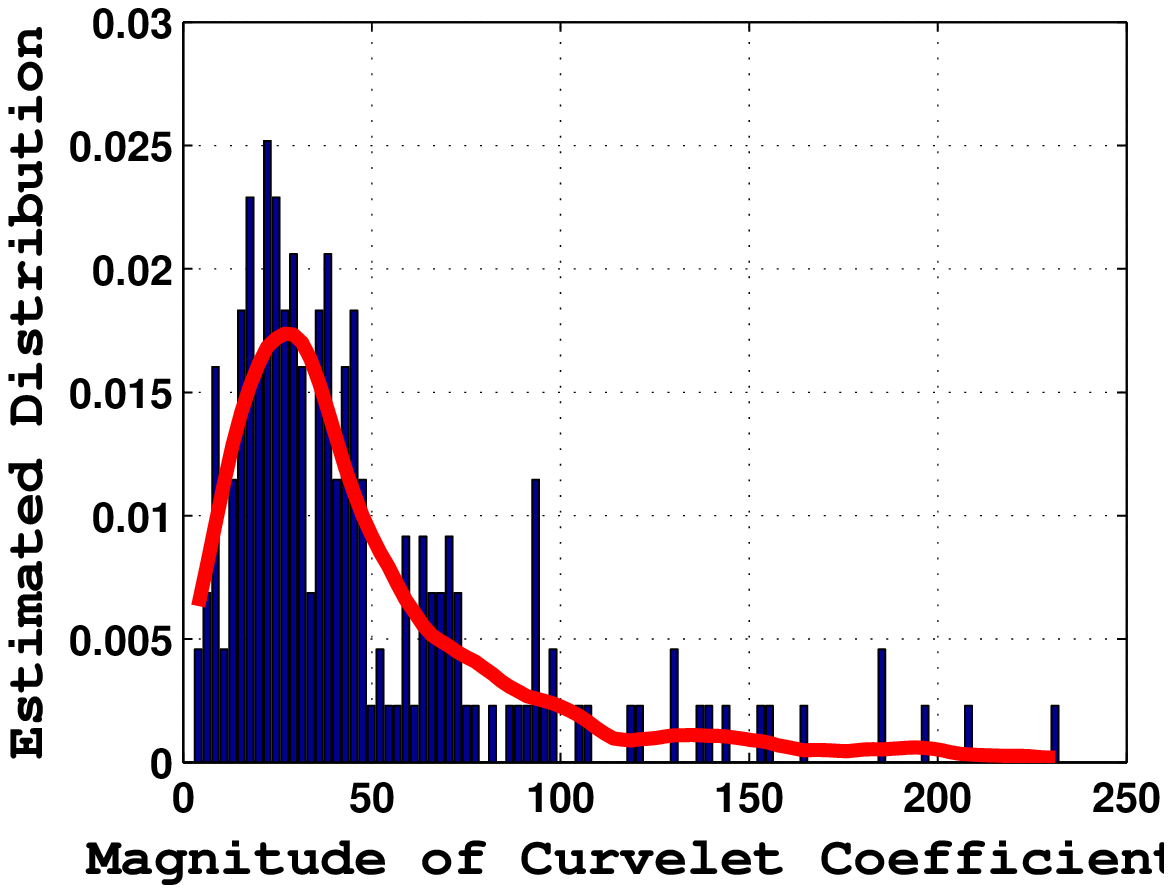}}
    \hspace{0.3cm}
    \subfloat[\label{PDF_CTsc3or1GaussN}]{\includegraphics[scale=0.51]{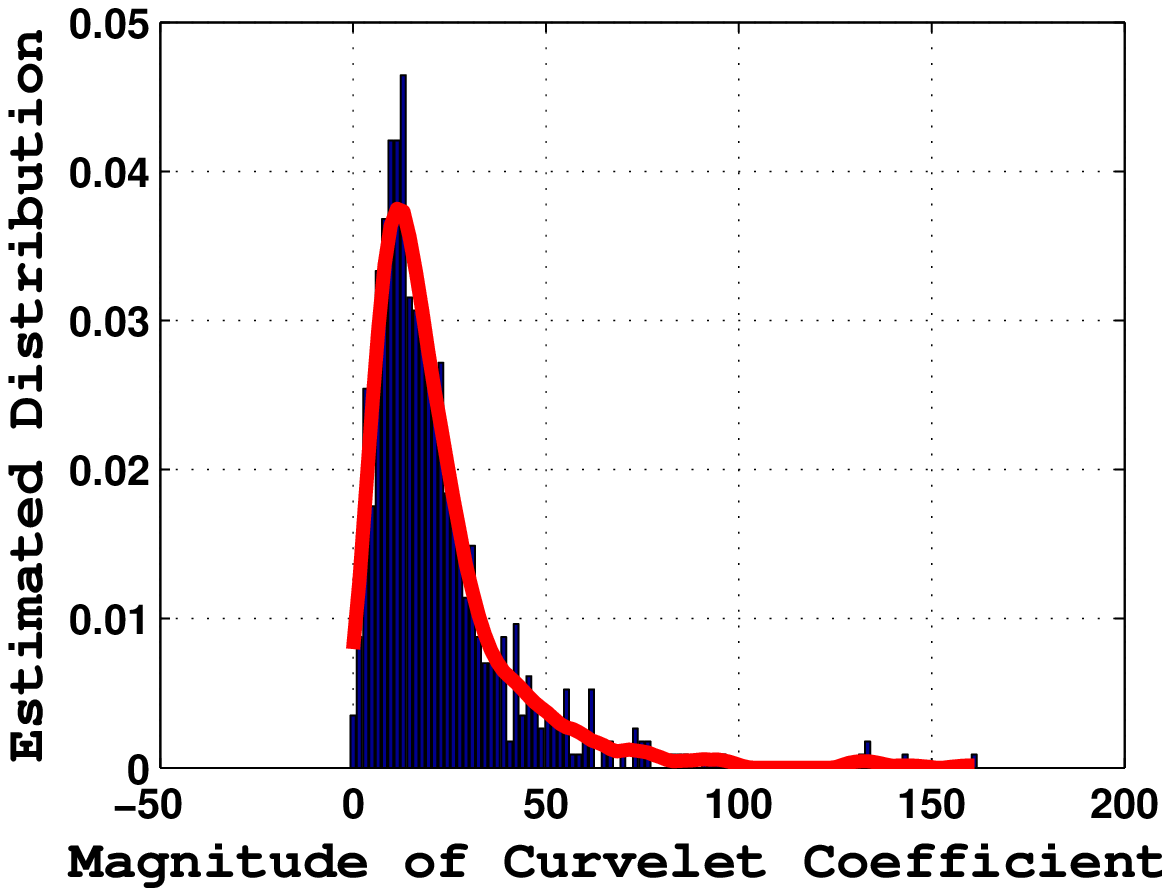}}\\
    \hspace{0.01cm}
    \subfloat[\label{PDF_CTsc4or1GaussN}]{\includegraphics[scale=0.51]{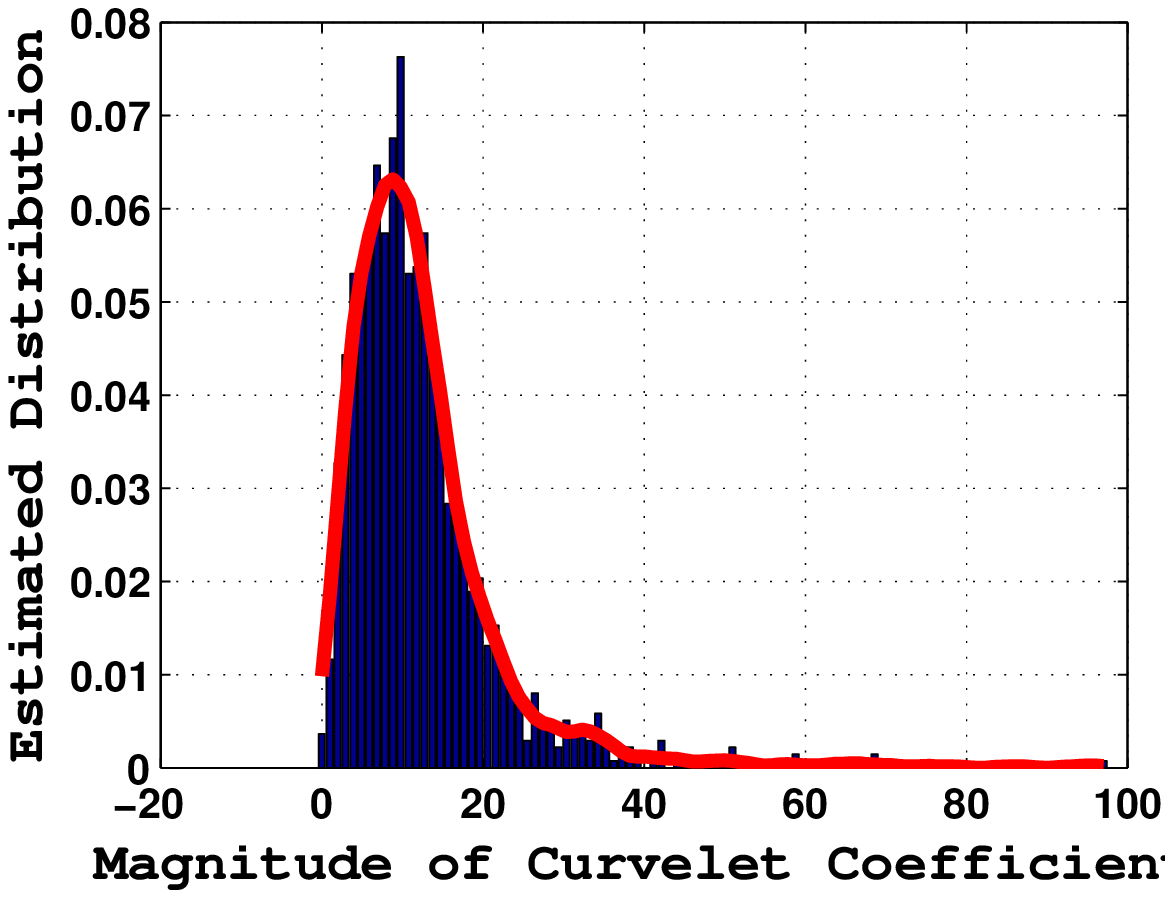}}
    \hspace{0.3cm}
    \subfloat[\label{PDF_CTsc5or1GaussN}]{\includegraphics[scale=0.51]{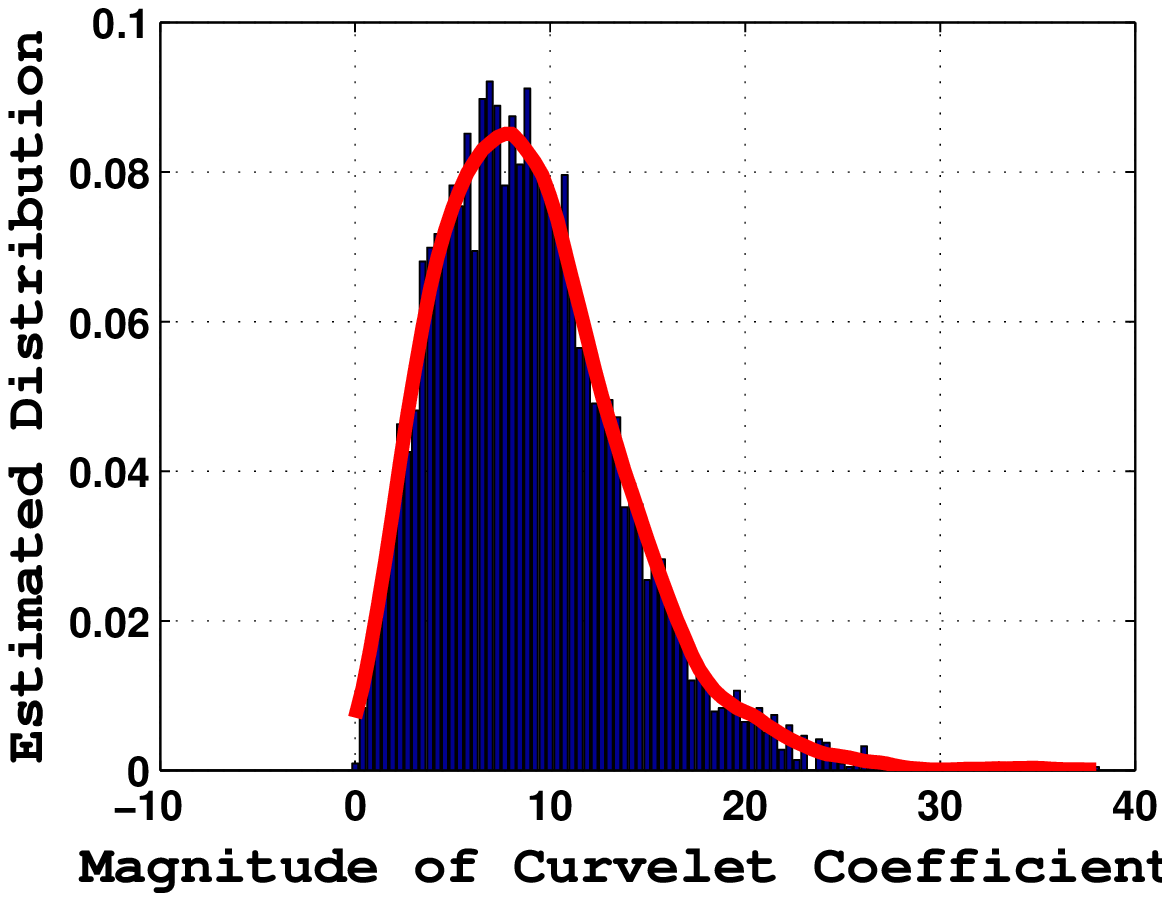}}
    \caption{\small{Estimated distribution (PDF) of noisy image (contaminated with AWGN $\sigma = 25$) Curvelet coefficients at (a) scale, $\gamma = 2$ \& orientation $o = 2$; (b) scale, $\gamma = 3$, orientation $o = 3$; (c) scale, $\gamma = 4$, orientation $o = 4$; (d) scale, $\gamma = 5$, orientation $o = 5$.}}
    \label{Fig:Sec2:Estimated_PDF_CT}
\end{figure*}

\section{Noise Analysis in Curvelet Domain}
\label{Sec2:NoiseDistribution}
Curvelet transform decomposes the pixel energy into magnitude and phase components. Literature indicates that the phases of any complex signal transform retain more structural information and are less sensitive to additive Gaussian noise compared to their magnitude \cite{skarbnik2009importance,panigrahi2013quantitative,panigrahi2016phases}. Assuming, that the Curvelet uses tight frames to represent any square integral function $f$ then it must obey Parseval's identity as \cite{ma2010curvelet}:

\begin{equation}\label{Eq:SubSec:CurveletSeriesApprox}
    f = \sum_{\gamma, \tau, o}{\langle f, \phi_{\gamma, \tau, o} \rangle}\phi_{\gamma, \tau, o}
\end{equation}

Panigrahi \emph{et al.} \cite{panigrahi2016phases} utilized the linearity property of Curvelet to derive the sensitivity of magnitude, $\frac{\partial|Y|}{{\partial}{|N|}} $ and phase, $\frac{\partial {\varphi_{Y}}}{{\partial}{|N|}}$ as:

\begin{equation}\label{Eq:Sec2:Sensitivity_Mag_WRT.Mag}
    \frac{\partial|Y|}{{\partial}{|N|}} = \frac{|N| + |Z|cos(\varphi_{Z} - \varphi_{N})}{\left(|Z|^2 + |N|^2 + 2|Z||N|cos(\varphi_{Z} - \varphi_{N})\right)^{1/2}}
\end{equation}

\begin{equation}\label{Eq:Sec2:Sensitivity_Phase_WRT.Mag}
    \frac{\partial {\varphi_{Y}}}{{\partial}{|N|}} = \frac{|Z|sin(\varphi_{N} - \varphi_{Z})}{|Z|^2 + |N|^2 + 2|Z||N|cos(\varphi_{Z} - \varphi_{N})}
\end{equation}

where, $\varphi_{Z}$ and $\varphi_{N}$ are the phase angles of the latent image, $z$ and noise, $\eta$ respectively. For experimental validation, we defined noise sensitivity as the ratio of difference between the magnitudes (or phase) of noisy curvelet coefficients to that of magnitude difference between the coefficients of AWGN for consecutive noise levels. The average of this measure is normalized between $[0,1]$ for $\sigma = [5,100]$. The sensitivity measure of curvelet magnitude and phase for each reference image of TID$2008$ database \cite{ponomarenko2009tid2008} contaminated with AWGN are shown in Fig.\ref{MagSensitivityPlot_AWGN} and \ref{PhaseSensitivityPlot_AWGN}, respectively. It indicates that the rate of change of Curvelet phase is less corrupted by AWGN compared to its magnitude. This justifies the preservation of phase in noise subspace, while estimating signal residuals using JBF.

The estimated PDF of Curvelet coefficients of a noisy image at different scales and orientations is shown in Fig.\ref{Fig:Sec2:Estimated_PDF_CT}.  It can be observed that the anisotropic scaling of Curvelet transform shapes the PDF of its magnitude at any scale ($\gamma$) and orientation ($o$) to leptokurtic i.e. very sharp peak at zero amplitude and extended tail on the either sides \cite{boubchir2005multivariate,rabbani2007image}. However, the threshold in \cite{starck2002curvelet} is estimated by assuming Gaussian distribution of the curvelet coefficients \cite{rabbani2007image}. The imprecise assumption in Curvelet thresholding leaves a few portions of the signal components in the noise subspace. We then re-estimate the thresholded coefficients to recover the suppressed signal in noise subspace using JBF. The current approach also aides in preserving the corresponding phase information that may strengthen the edges of denoised image.

\tikzstyle{box} = [draw, rectangle,thick, fill=gray!20, text width=3em, text centered, minimum height = 12mm]
\tikzstyle{box1} = [draw, rectangle,thick, fill=yellow!10, rounded corners, text width=4em, text centered, minimum height = 12mm]
\tikzstyle{box2} = [draw, rectangle,thick, fill=blue!10, rounded corners, text width=5em, text centered, minimum height = 12mm]
\tikzstyle{box3} = [draw, rectangle,thick, fill=green!10, rounded corners, text width=2.5em, text centered, minimum height = 12mm]
\tikzstyle{sum} = [draw, fill=blue!20, circle, text width=1em, node distance=1cm]
\tikzstyle{dot} = [draw, circle,fill=black,inner sep=0pt,minimum size=6pt, node distance=0.5cm]
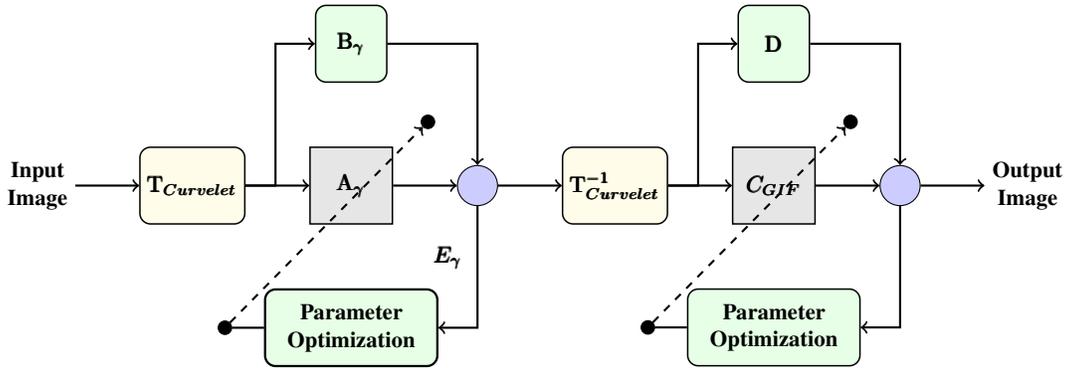
\begin{figure*}[!ht]
\begin{center}
\begin{tikzpicture}[auto,scale=0.85, transform shape]
\node[box1, line width=0.2mm] (a) {$\pmb{\mathrm{T}_{Curvelet}}$};
\node[left=of a,font=\bfseries,align=center] (aux1) {Input \\ Image};
\node[box,line width=0.2mm,right=of a] (b) {$\pmb{\mathrm{A}_{\gamma}}$};
\node[box3,line width=0.2mm,align=center,above=of b] (Expt) {$\pmb{\mathrm{B}_{\gamma}}$};
\node[dot,below=of Expt, xshift=1.2cm] (Dot0) {};
\node[box3,line width=0.3mm,font=\bfseries,text width=7em,below=of b] (Par) {Parameter Optimization};
\node[dot,font=\bfseries,left=of Par] (Dot1) {};
\node[sum, right=of b,font=\bfseries,align=center] (error) {};
\node[box1,line width=0.2mm,right=of error] (Inv) {$\pmb{\mathrm{T}^{-1}_{Curvelet}}$};
\node[box,line width=0.2mm,right=of Inv] (GIF) {$\pmb{C_{GIF}}$};
\node[box3,line width=0.2mm,above=of GIF] (Det) {$\pmb{\mathrm{D}}$};
\node[dot,below=of Det, xshift=1.2cm] (Dot2) {};
\node[box3,line width=0.2mm,font=\bfseries,text width=7em,below=of GIF] (ParD) {Parameter Optimization};
\node[dot,font=\bfseries,left=of ParD] (Dot3) {};
\node[sum,align=center,right=of GIF] (error1) {};
\node[right=of error1,font=\bfseries,align=center] (Out) {Output \\ Image};
\draw[->,black,thick] (aux1) -- (a);
\draw[->,black,thick] (a) -- (b);
\draw[->,black,thick] (b) -- (error);
\draw[->,black,thick] (error) -- (Inv);
\draw [->,black,thick] (a) -- ++ (1.3,0) |-  (Expt);
\draw [->,black,thick] (Expt) -| node[pos=0.99] {}
        node [near end] {} (error);
\draw [->,black,thick] (error) |- node [yshift=4em, xshift=-1.3em] {$\pmb{E_{\gamma}}$} (Par);
\draw [-,black,thick] (Par) -- (Dot1);
\draw [-,black,thick] (ParD) -- (Dot3);
\draw [->,black,dashed,thick] (Dot1) -- (Dot0);
\draw [->,black,dashed,thick] (Dot3) -- (Dot2);
\draw[->,black,thick] (Inv) -- (GIF);
\draw [->,black,thick] (Inv) -- ++ (1.3,0) |-  (Det);
\draw [->,black,thick] (Det) -| node[pos=0.99] {}
        node [near end] {} (error1);
\draw [->,black,thick] (error1) |- node[pos=0.99] {}
        node [near end] {} (ParD);
\draw[->,black,thick] (error1) -- (Out);
\draw[->,black,thick] (GIF) -- (error1);
\end{tikzpicture}
\end{center}
\caption{Solution Approach to the Proposed Problem Definition.}
\label{Ch1:Fig:Sec:ProblemDefinationProp}
\end{figure*}

\section{Proposed Image Denoising Framework}
\label{Sec3:ProposedMethod}
In contrast to the Curvelet thresholding (CT) approach in Eq.\ref{Eq:Sec1:ThreshDenoising}, the proposed formulation in Eq.\ref{Eq:Sec:CT_ProbDef_JBF} preserves the signal components in the coarser scales, if the coefficients, $C \ge \lambda_{\gamma}$. The spilled or residual signals in the noise subspace (for $C < \lambda_{\gamma}$) are estimated using multiscale Joint Bilateral Filter (JBF), $\pmb{\mathrm{A}_{JBF, C < \lambda_{\gamma}}}$. On the other hand the high frequency components such as: edges, textures and small details in the finest scale are retained maximally using Bilateral Filter, $\pmb{\mathrm{A}_{BF, \gamma}}$. In the proposed formulation the parameters (of different filters in Curvelet domain) are optimized by minimizing the mean square error (MSE) $\pmb{\|E_\gamma\|}$ between the estimated and the desired coefficients, assuming obtained by using the operator, $\pmb{\mathrm{B}_{\gamma}}$. Further, the post-processing Guided Image Filter (GIF), $\pmb{\mathrm{C}_{GIF}}$ in the spatial domain is adopted for better localization and preservation of local image structures like: edges, textures and small details.

\begin{equation}\label{Eq:Sec:CT_ProbDef_JBF}
    \centering
    \pmb{\mathrm{A}_{\gamma}} = \left[\pmb{\overline{\mathbf{I}}_{C \ge \lambda_{\gamma}}} \mid \pmb{\mathrm{A}_{JBF, C < \lambda_{\gamma}}} \mid \pmb{\mathrm{A}_{BF, \gamma}} \right]
\end{equation}

The various steps involved in the  proposed multiscale framework is illustrated in Fig.\ref{Fig:Sec3:BD_ProposedMethod}. The JBF followed by hard thresholding in the coarser scales is used to recover the lost signal in noise subspace. In the finest scale, BF ensures the preservation of well connected edges. However, the inevitable ringing artifacts in the reconstructed image requires further processing. We applied guided image filter (GIF) in the reconstructed image to localize the edges and to preserve the small details and textures of the latent image, maximally. The complete description and formulation involved in each step of the proposed image denoising technique are discussed in the following subsections.


\tikzstyle{block} = [draw, rectangle,thick, fill=blue!10, text width=6em, text centered, minimum height = 12mm]
\tikzstyle{block1} = [draw, rectangle,thick, fill=gray!30, text width=5.8em, text centered, minimum height = 12mm]
\tikzstyle{block2} = [draw, rectangle,thick, fill=green!10, text width=10em, text centered, minimum height = 10mm]
\tikzstyle{block3} = [draw, rectangle,thick, fill=yellow!20, text width=7em, text centered, minimum height = 12mm]
\tikzstyle{block4} = [draw, rectangle,thick, fill=magenta!20, text width=10em, text centered, minimum height = 12mm]
\tikzstyle{block5} = [draw, rectangle,thick, fill=magenta!10, text width=7em, text centered, minimum height = 12mm]
\tikzstyle{line} = [draw, -stealth, thick]

\begin{figure*}[!ht]
\hspace*{-1.5em}\raisebox{0em}{\begin{tikzpicture}[auto,scale=0.85, transform shape]
  \node [block] (NI) {\footnotesize{Noisy Image \\ $y(\cdot)$}};
  \node [block3,right of=NI, xshift=5.4em](CT){\footnotesize{Curvelet Transform \\ $\textbf{T}_{\gamma}(\cdot)$}};
  \node [block4,below of=CT,yshift=-5em, xshift=14em](BF_F){\footnotesize{Preservation of High Frequency Components}};
  \node [yshift=-1.8ex, black] at (BF_F.south) {\textbf{\footnotesize{Bilateral Filter}}};
  \node [yshift=-5.5ex, black] at (BF_F.south) {\pmb{$\sigma_{d}^{f}$, $\sigma_{r}^{f}$}};
  \node [block1,above of=CT,yshift=4.5em, xshift=9em](BF_C){\footnotesize{Noisy Coeff. in Different: $\gamma$ \& $o$}};
  \node [block1,above of=BF_C,yshift=1em, xshift=10em](Ab_Th){\footnotesize{Thresholded Coefficients}};
  \node [yshift=-1.5ex, black] at (Ab_Th.south) {\textbf{\footnotesize{Initial Estimate}}};
  \node [block1,below of=BF_C,yshift=-3em, xshift=15em](Bl_Th){\footnotesize{Estimation of Signal in Noise Subspace}};
  \node [yshift=-1.8ex, xshift=-1.5ex, black] at (Bl_Th.south) {\textbf{\footnotesize{Joint Bilateral Filter}}};
  \node [yshift=-5.5ex, black] at (Bl_Th.south) {\pmb{($\sigma_{d}^{c}$, $\alpha$)}};
  \node [block2,right of=CT,xshift=28em,yshift=1.5em,rotate=90](Est_Coeff){\footnotesize{Denoised Coefficients in Different Scales ($\gamma$)}};
  \node [block3,right of=Est_Coeff, xshift=3.8em](ICT){\footnotesize{Inverse Curvelet Transform \\ $\textbf{T}_{\gamma}^{-1}(\cdot)$}};
  \node [block5,right of=ICT,xshift=6em](GIF){\footnotesize{Guided Image Filter \\ $\epsilon = k_1 \times \sigma$}};
  \node [block,below of=GIF,xshift=-0em,yshift=-3.5em](DI){\footnotesize{Denoised Image \\ $\hat{z}(\cdot)$}};
  \draw [line] (NI) -- (CT);
  \draw [line] (CT) -- ++ (1.65,0) |-  (BF_C);
  \draw [line] (CT) -- ++ (1.65,0) |-  (BF_F);
  \draw [line] (BF_C) -- ++ (1.65,0) |- node[yshift=0.2em, xshift=0.5em] {\textbf{\footnotesize{$\geq {\lambda_{\gamma, o}}$}}} (Ab_Th);
  \draw [line][text=blue] (BF_C) -- ++ (1.65,0) |- node[yshift=-2.5em, xshift=1em] {\textbf{\footnotesize{\begin{tabular}{cc}
Photometric & \\ Distance &   \end{tabular}}}} (Bl_Th);
  \draw [line] (BF_F.east) -|  (Est_Coeff.west);
  \draw [line] (Ab_Th.east) -|  (Bl_Th.north);
  \draw [line] (Bl_Th.east) --  (Est_Coeff.north);
  \draw [line] (Est_Coeff.south) --  (ICT);
  \draw [line] (ICT) -- (GIF);
 \draw [line] (GIF.east) -- ++ (0.4,0) |-  (DI.east);
 \node (X) [draw=blue, very thick, dotted, fit= (BF_F), inner sep=0.6cm] {};
 \node [xshift=-2ex, black, rotate=90] at (X.west) {\textbf{The Finest Scale}};
 \node (Y) [draw=blue, very thick, dotted, fit= (BF_C)(Ab_Th)(Bl_Th), inner sep=0.9cm] {};
 \node [xshift=-2ex, black, rotate=90] at (Y.west) {\textbf{Coarser Scales}};
\end{tikzpicture}}
\caption{\small{Illustration of various steps involved in the proposed image denoising framework.}}
\label{Fig:Sec3:BD_ProposedMethod}
\end{figure*}
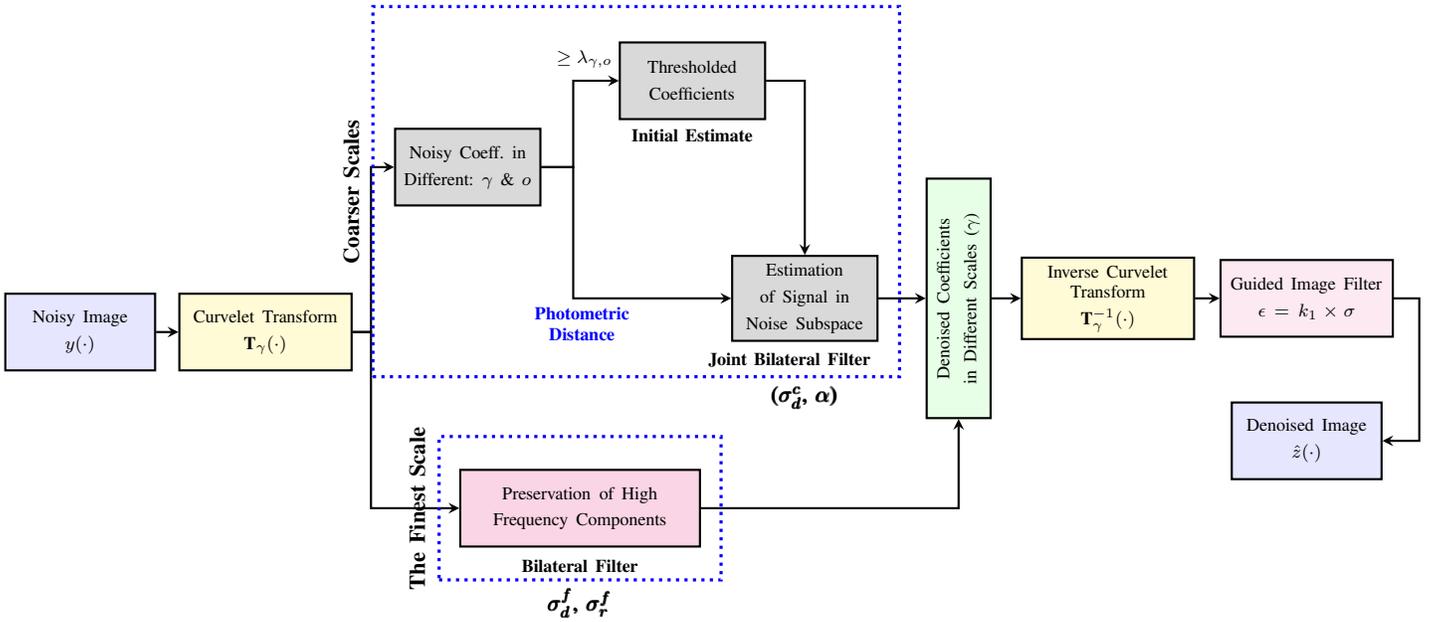

\subsection{Recovering Signal from Noise Subspace using JBF}
\label{SubSec3:CoarserScales}
Redefining Eq.\ref{Eq:Sec1:Hard_Threshold}, for curvelet transform and assuming $Y_{\gamma,o}$ as the noisy coefficients at any (coarser) scale $\gamma$ and orientation $o$, the thresholded coefficient is obtained as \cite{starck2002curvelet}:

\begin{table*}[ht]
\centering
\small
\caption{\small{Average Correlation Coefficient between the Estimated coefficient in noise subspace ---using both Curvelet hard thresholding (CT(Hard)) \cite{starck2002curvelet} and Proposed method--- and Original Curvelet Coefficient at different Coarser Scales on TID$2008$ Image Database \cite{ponomarenko2009tid2008}.}}
\label{Table:Sec3_Correlation_Coefficient}
\begin{tabular}{|c|cc|cc|cc|cc|}
\hline
\multicolumn{1}{|c|}{\multirow{2}{*}{\begin{tabular}[c]{@{}c@{}}Noise \\ Strength\end{tabular}}} & \multicolumn{2}{c|}{$\gamma = 2$}                              & \multicolumn{2}{c|}{$\gamma = 3$}                              & \multicolumn{2}{c|}{$\gamma = 4$}                              & \multicolumn{2}{c|}{$\gamma =5$}                               \\ \cline{2-9}
\multicolumn{1}{|c|}{}                                                                           & \multicolumn{1}{c|}{CT (Hard)} & \multicolumn{1}{c|}{Proposed} & \multicolumn{1}{c|}{CT (Hard)} & \multicolumn{1}{c|}{Proposed} & \multicolumn{1}{c|}{CT (Hard)} & \multicolumn{1}{c|}{Proposed} & \multicolumn{1}{c|}{CT (Hard)} & \multicolumn{1}{c|}{Proposed} \\ \hline \hline
$\sigma = 10$        & 0.9932           & 0.9956          & 0.9439        & 0.9798        & 0.7910       & 0.9249      & 0.4681       & 0.7396\\
$\sigma = 20$        & 0.9737           & 0.9877          & 0.8508        & 0.9476        & 0.5723       & 0.8066      & 0.2388       & 0.5366\\
$\sigma = 25$        & 0.9609           & 0.9815          & 0.7985        & 0.9257        & 0.4835       & 0.7463      & 0.1755       & 0.4629\\
$\sigma = 30$        & 0.9463           & 0.9741          & 0.7454        & 0.9017        & 0.4071       & 0.6881      & 0.1320       & 0.4027\\
$\sigma = 40$        & 0.9130           & 0.9536          & 0.6479        & 0.8541        & 0.2916       & 0.6136      & 0.0799       & 0.3494 \\
$\sigma = 50$        & 0.8737           & 0.9370          & 0.5600        & 0.8105        & 0.2162       & 0.5310      & 0.0520       & 0.2986 \\
$\sigma = 75$        & 0.7636           & 0.8839          & 0.3892        & 0.6963        & 0.1177       & 0.3795      & 0.0231       & 0.2245 \\
\hline
\end{tabular}
\end{table*}


\begin{equation}\label{Eq:Sec3:CT_Hard_Threshold}
    \centering
 Y_{\gamma,o}^{\prime}=\begin{cases}
    Y_{\gamma,o}~, & \text{if\, $|Y_{\gamma,o}| > {\lambda_\gamma,o}$}\\
    0~, & \text{otherwise}
 \end{cases}
\end{equation}

where, the threshold, $\lambda_{\gamma,o}$ is defined as:

\begin{equation}\label{Eq:Sec3:Threshold}
    \centering
    T_{\gamma,o} = {k}{\sigma}{\sigma_{\gamma,o}}
\end{equation}

Here, $k$ is assumed to be a scale dependant constant and $\sigma_{\gamma,o}$ is estimated using Monte Carlo simulation of the Curvelet transform of a few standard white noise images \cite{starck2002curvelet}.

%

The initial estimate $Y_{\gamma,o}^{\prime}$ introduces ringing artifacts and looses phase information due to thresholding. The thresholded coefficients in Eq.\ref{Eq:Sec3:CT_Hard_Threshold} is further processed using JBF \cite{petschnigg2004digital} to recover the lost signal from the noise subspace (below threshold) as:

\begin{equation}
\label{Eq:Sec3:JBF_CoarserScales}
  \tilde{Y}_{\gamma,o}(U) = \frac{1}{C^c} \sum_{r \in \aleph_{JBF}(U)}e^{-\frac{{\|r - U\|}^2}{2\left(\sigma_{d}^{c}\right)^2}} e^{-\frac{{\|Y_{\gamma,o}(r) - Y_{\gamma,o}(U)\|}^2}{2\left(\sigma_{r}^{c}\right)^2}}Y_{\gamma,o}^{\prime}(r)
\end{equation}

where, $\aleph_{JBF}(U)$ is the spatial neighboring window (in Curvelet domain) around $U$. The normalizing constant $C^c$ ensures the coefficient weights sum must converge to $1.0$. The parameters, $\sigma_{d}^{c}$ and $\sigma_{r}^{c}$ are the standard deviations of the exponential functions defined for spatial and photometric distances, respectively. The photometric distance parameter is selected as $\alpha$ times of the total range of coefficient magnitude.

The justification of JBF to extract information from below threshold is empirically examined over the Curvelet hard thresholding technique \cite{starck2002curvelet}. The original Curvelet coefficient and estimated coefficients using both Curvelet hard thresholding and proposed method in coarser scales are calculated for different noise strength $\sigma$ on TID2008 image database. Table.\ref{Table:Sec3_Correlation_Coefficient} indicates that the estimated coefficients in noise subspace are highly correlated with original Curvelet coefficient. Therefore the proposed estimation of signal in noise subspace suppresses noise and retains better interdependency among the coefficients at different scales. Since the phase is indispensable for image representation \cite{panigrahi2013quantitative} and it is also less corrupted compared to the Curvelet magnitude, thus we retain the corresponding phase of estimated magnitude in noise subspace.

%

\begin{figure*}[!ht]
    \centering
    \subfloat[\label{Original_FineScale}]{\includegraphics[scale=0.43]{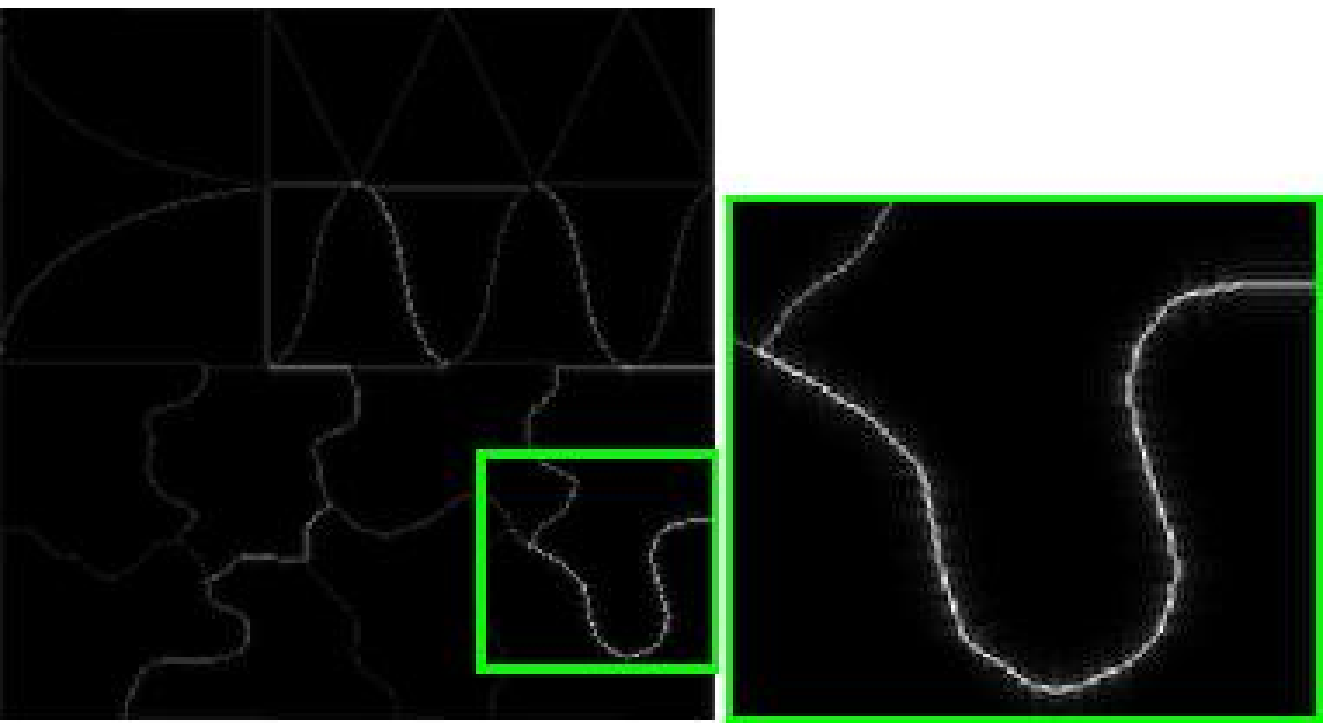}}
    \hspace{0.1cm}
    \subfloat[\label{CTu_FineScale_DeCoeff}]{\includegraphics[scale=0.43]{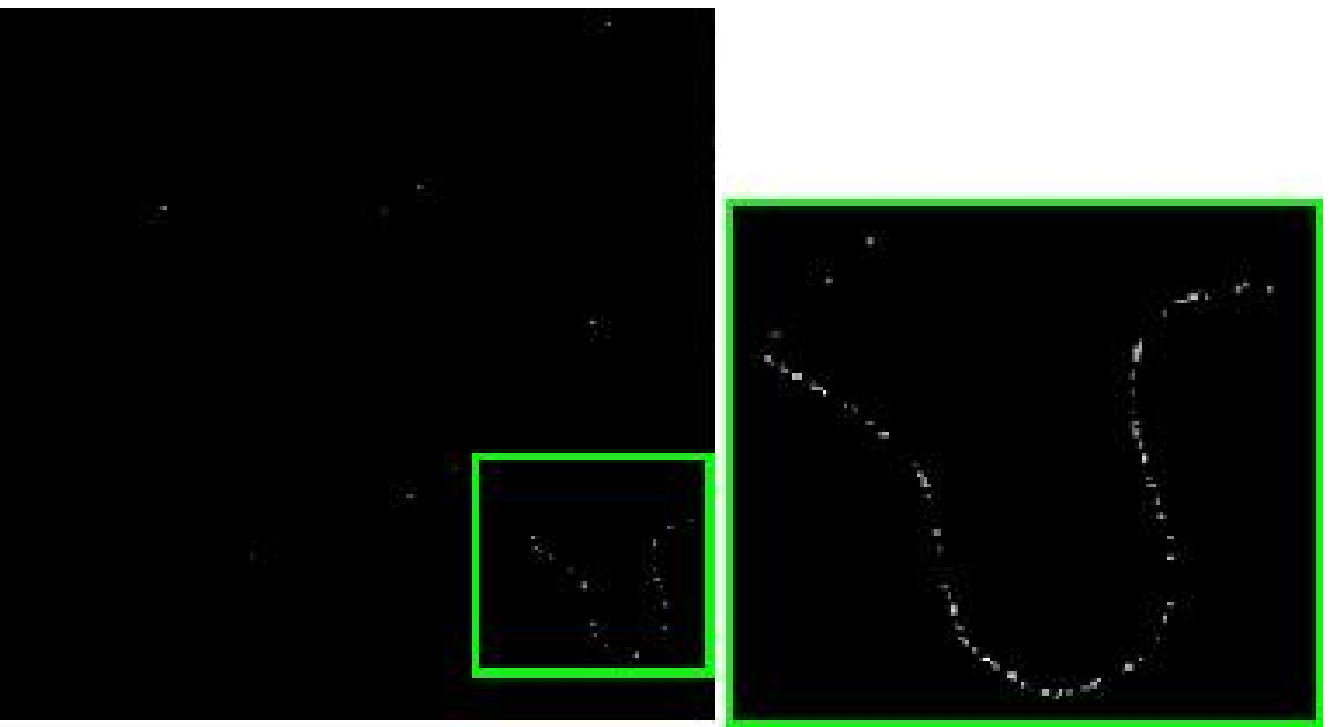}}
    \hspace{0.1cm}
    \subfloat[\label{Proposed_FineScale_DeCoeff}]{\includegraphics[scale=0.43]{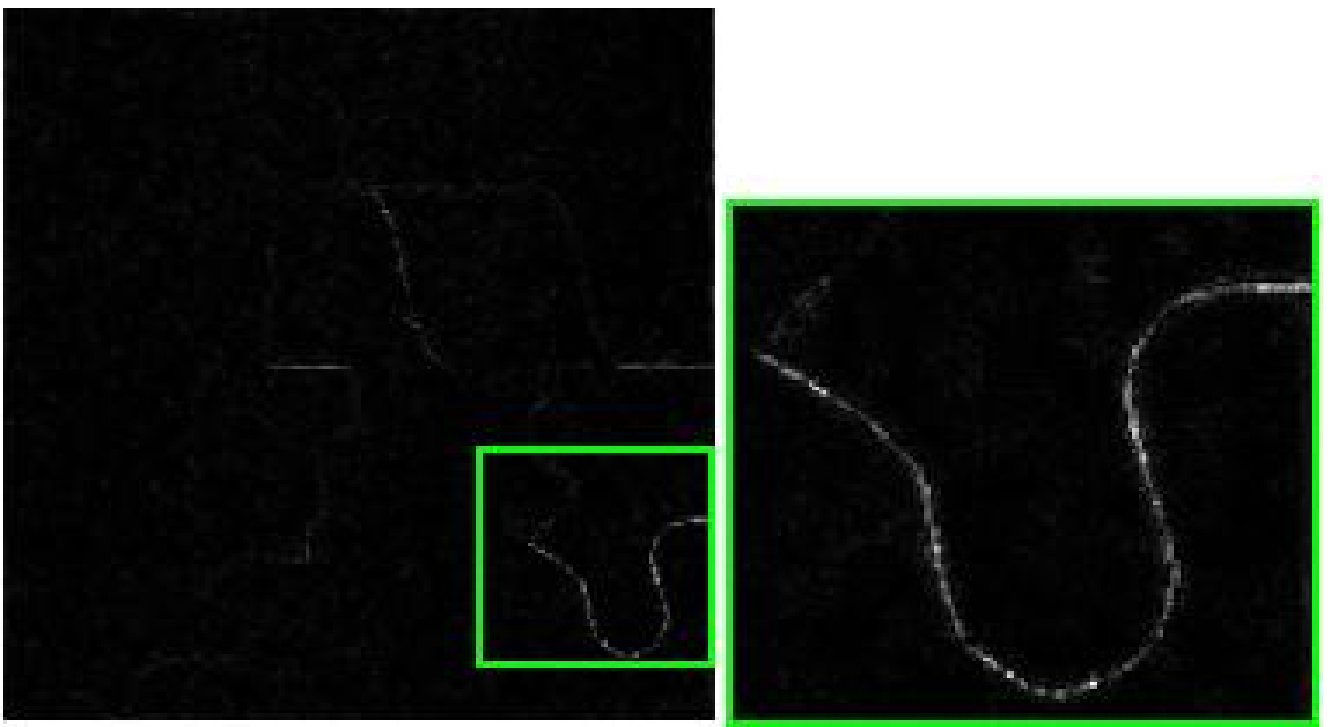}}
    \caption{\small{Curvelet coefficients in the fine scale for (a) Original Texture image and the recovered details from the AWGN of $\sigma=25$ using (b) Curvelet Thresholding and (c) Bilateral filtering.}}
    \label{Fig:Sec3.2:Curvelet_FineScale}
\end{figure*}

\subsection{Retaining Small Details using BF}
\label{SubSec3:FinerScale}
High frequency components such as: edges, textures and small details are captured by the coefficients in the finest scale. It is observed that few signal and noise magnitudes in this scale are comparable. The application of hard threshold removes both noise and a part of these details resulting in granular effect -- due to discontinuous edges as shown in Fig.\ref{CTu_FineScale_DeCoeff} -- in the restored image. Thus, Unlike \cite{starck2002curvelet}, we have computed the denoised coefficients in the finest scale, $\gamma$ using BF \cite{tomasi1998bilateral}:

\begin{equation}
\label{Eq:Sec3:BF_FinerScales}
  \tilde{Y}_{\gamma}(U) = \frac{1}{C^f} \sum_{r \in \aleph_{BF}(U)}e^{-\frac{{\|r - U\|}^2}{2\left(\sigma_{d}^{f}\right)^2}} e^{-\frac{{\|Y_{\gamma}(r) - Y_{\gamma}(U)\|}^2}{2\left(\sigma_{r}^{f}\right)^2}}{Y}_{\gamma}(r)
\end{equation}

Similar to the Eq.\ref{Eq:Sec3:JBF_CoarserScales}, the falling rate parameters of the BF kernels in spatial and intensity domain is defined as $\sigma_{d}^{f}$ and $\sigma_{r}^{f} = k_{r} \times \lambda_{\gamma}$, respectively. As illustrated in Fig.\ref{Fig:Sec3.2:Curvelet_FineScale} the filtered output in the finest scale removes the granular effect, while preserving the well-connected edges compared to thresholding \cite{starck2002curvelet}.

\subsection{Post Processing using GIF}
\label{subsec3.3:GIF_PostProcess}
Though the estimation of signal -- in coarser scale -- below threshold attenuates the sudden jumps in coefficient magnitude, still requires further processing to remove the ringing artifacts in the reconstructed image. The fast ($\mathcal{O}(M)$) ``Edge-Aware" Guided Image Filter (GIF) is considered here to suppress the distortions around the edges of the reconstructed image, $\tilde{z}$. In the absence of guidance image $G$, it is assumed that $G \equiv \tilde{z}$ and the GIF is reformulated as \cite{he2013guided}:

\begin{equation}\label{Eq:subsec:GIF_Final}
\hat{z}(\mathcal{P}) = \bar{a}_\mathcal{P} \tilde{z}(\mathcal{P}) + \bar{b}_\mathcal{P}
\end{equation}

where, \\ $\bar{a}_\mathcal{P} = \frac{1}{|\aleph_{GIF}(\mathcal{P}^\prime)|}\sum_{{\mathcal{P}^\prime} \in \aleph_{GIF}(\mathcal{P})} a_{\mathcal{P}^\prime}$ \\
\& $\bar{b}_\mathcal{P} = \frac{1}{|\aleph_{GIF}(\mathcal{P}^\prime)|}\sum_{{\mathcal{P}^\prime} \in \aleph_{GIF}(\mathcal{P})} b_{\mathcal{P}^\prime}$ are the average values of $a_{\mathcal{P}^\prime}$ and $b_{\mathcal{P}^\prime}$ for all the overlapping windows that covers the pixel $\mathcal{P}$. The linear parameters, $a_{\mathcal{P}^\prime}$ and $b_{\mathcal{P}^\prime}$ that ensures the preservation edges are estimated as:

\begin{subequations}\label{Eq:sec:GIF_SingleIm}
     \begin{align}
     a_{\mathcal{P}^\prime} &= \frac{\sigma_{\tilde{z}}^{2}(\mathcal{P}^\prime)}{\sigma_{\tilde{z}}^{2}(\mathcal{P}^\prime) + \epsilon}\label{Eq:sec:GIF_SingleIm1}\\
b_{\mathcal{P}^\prime} &= (1 - {a}_{\mathcal{P}^\prime}){\mu}_{\tilde{z}}(\mathcal{P}^\prime)\label{Eq:sec:GIF_SingleIm2}
     \end{align}
\end{subequations}

\begin{figure}[h]
    \centering
    \subfloat[\label{Fig:PPCT_BefGIF_Lena}]{\includegraphics[scale=0.38]{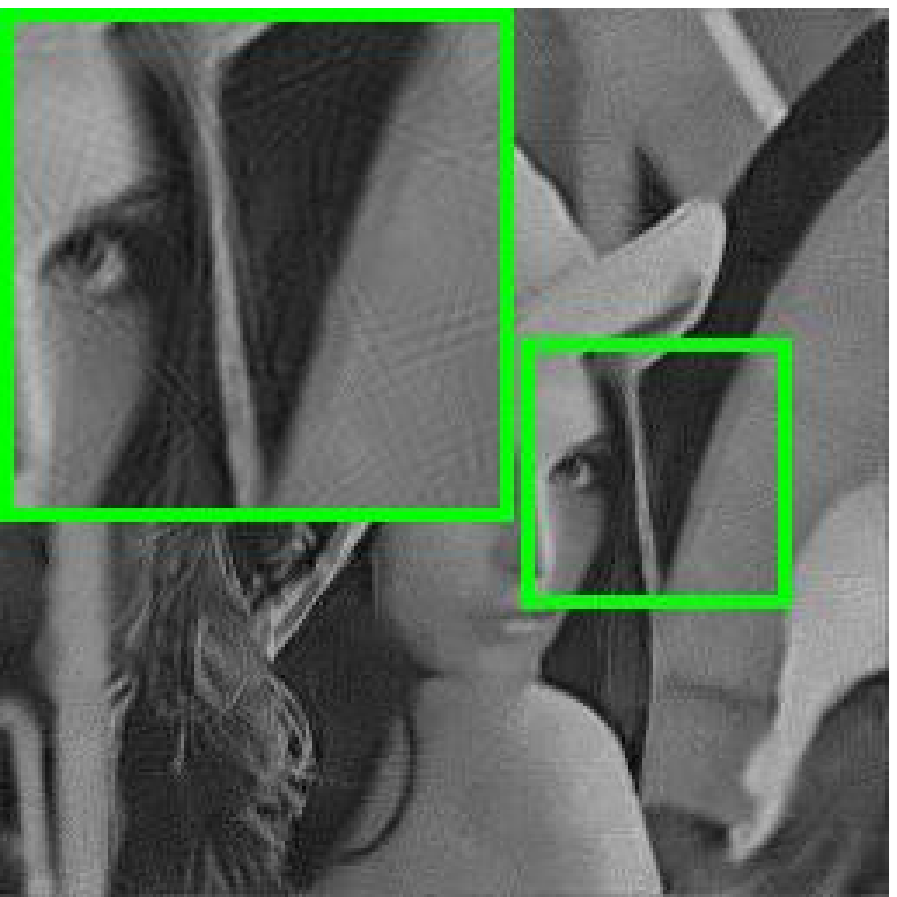}}
    \hspace{0.5cm}
    \subfloat[\label{Fig:PPCT_AftGIF_Lena}]{\includegraphics[scale=0.38]{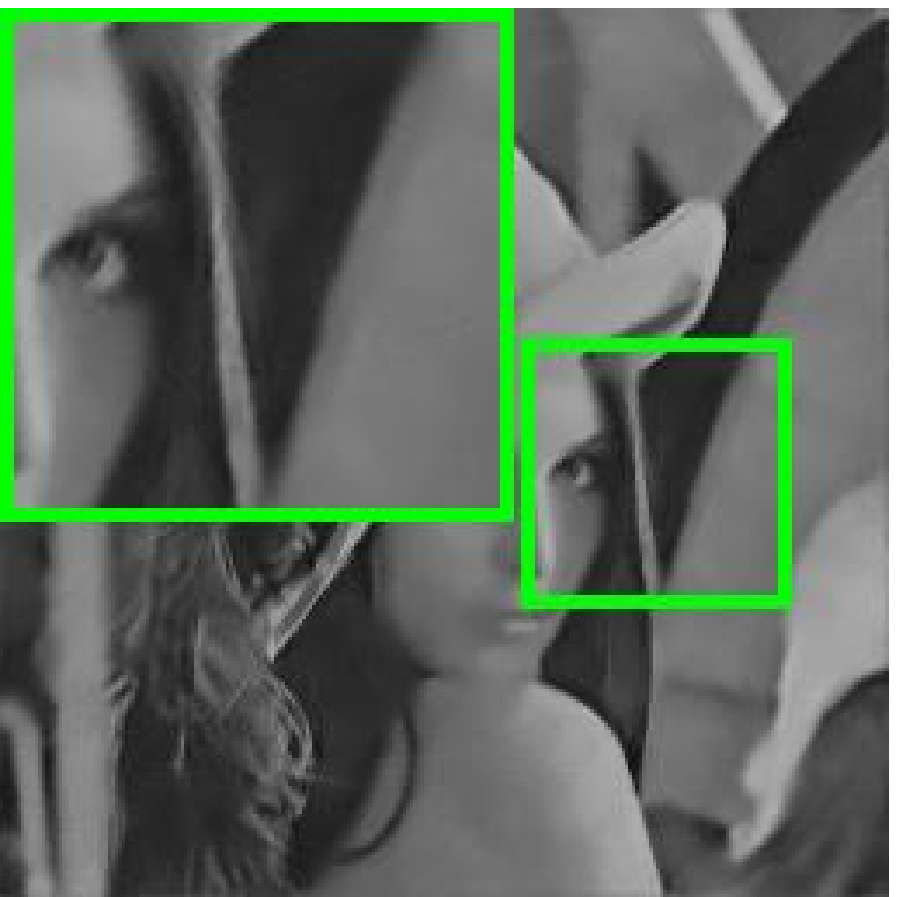}}
     \caption{\small{Proposed denoised method for Lena image ($\sigma$ = 40), (a) Before (PSNR = 29.4091, SSIM = 0.8254) and (b) After (PSNR = 30.001, SSIM = 0.8620) application of GIF.}}
    \label{Fig:subsecAlgoGray:GIF}
\end{figure}

The regularization parameter $\epsilon = k_1 \times \sigma$ penalizes the over smoothing of edges. The significance of the application of GIF can be observed in Fig.\ref{Fig:subsecAlgoGray:GIF}. Both the visual and quantitative improvement in terms of PSNR and SSIM \cite{wang2004image} measure demonstrate its efficacy in preserving small details, while suppressing the distortion around the edge.


\begin{figure*}[!ht]
    \centering
    \subfloat[\label{fig:PSNR_SigmaDR_TID10}]{\includegraphics[scale=0.6]{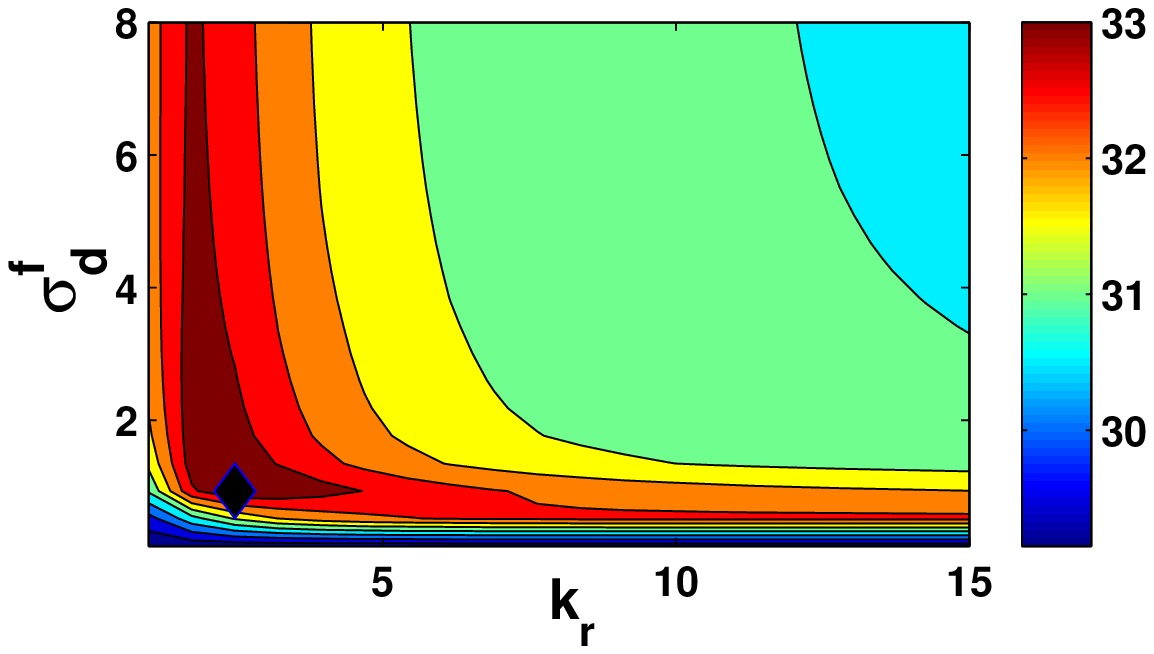}}
    \hspace{0.01cm}
    \subfloat[\label{fig:PSNR_SigmaDR_TID50}]{\includegraphics[scale=0.6]{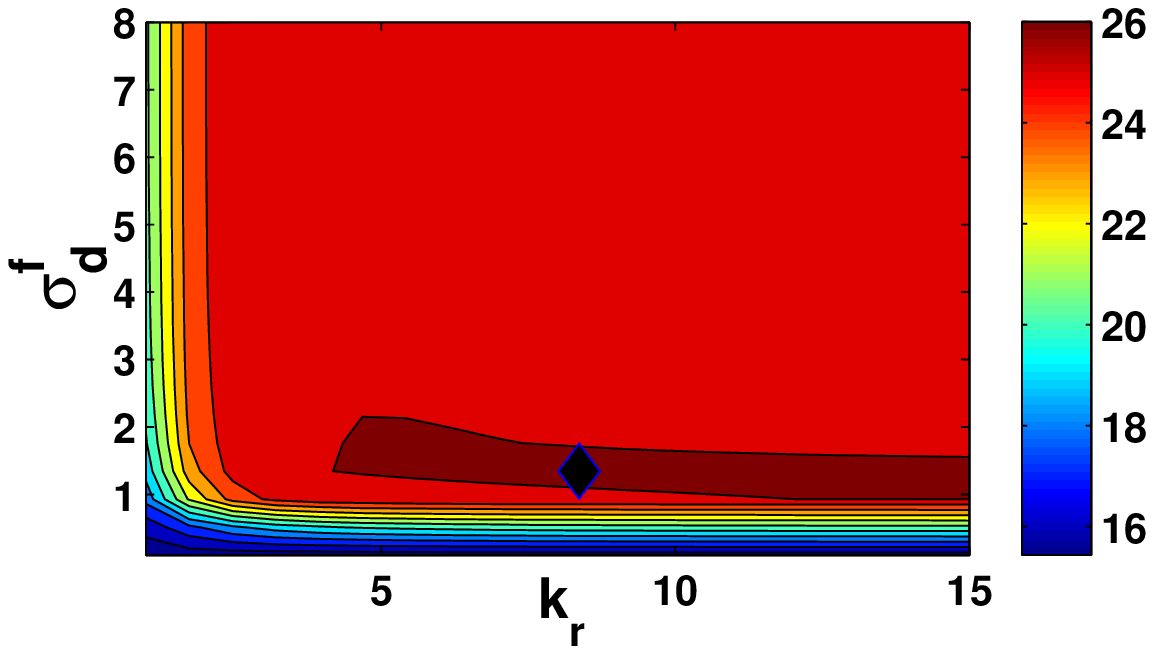}}\\
    \hspace{0.11cm}
    \subfloat[\label{fig:SSIM_SigmaDR_TID10}]{\includegraphics[scale=0.6]{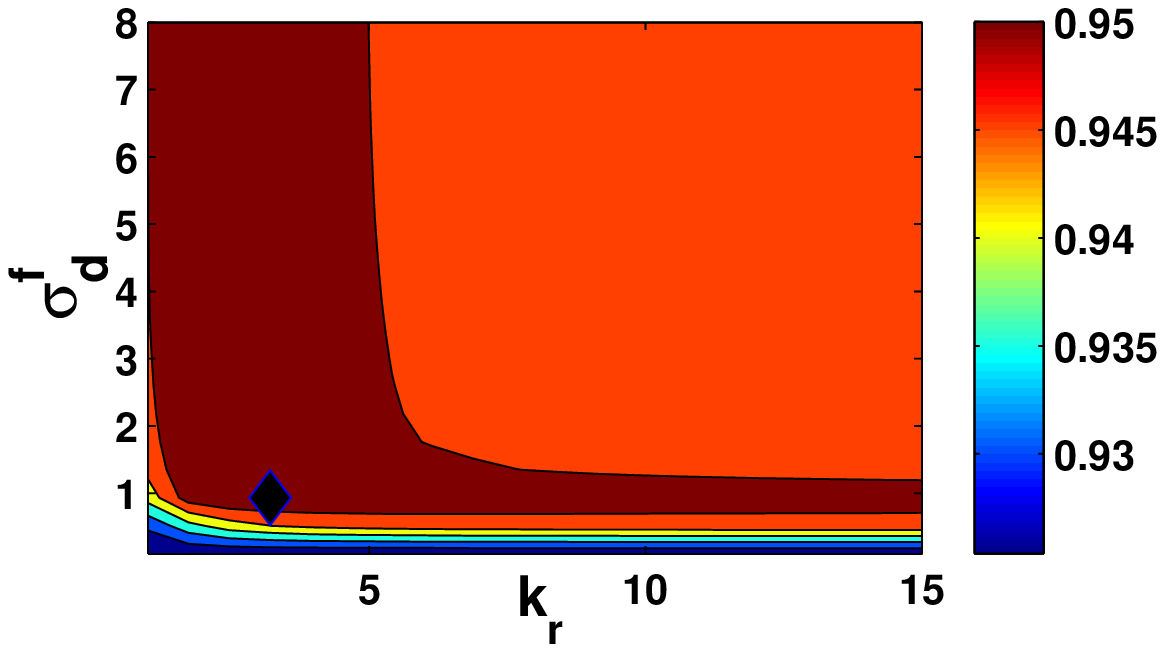}}
    \hspace{0.01cm}
   \subfloat[\label{fig:SSIM_SigmaDR_TID50}]{\includegraphics[scale=0.6]{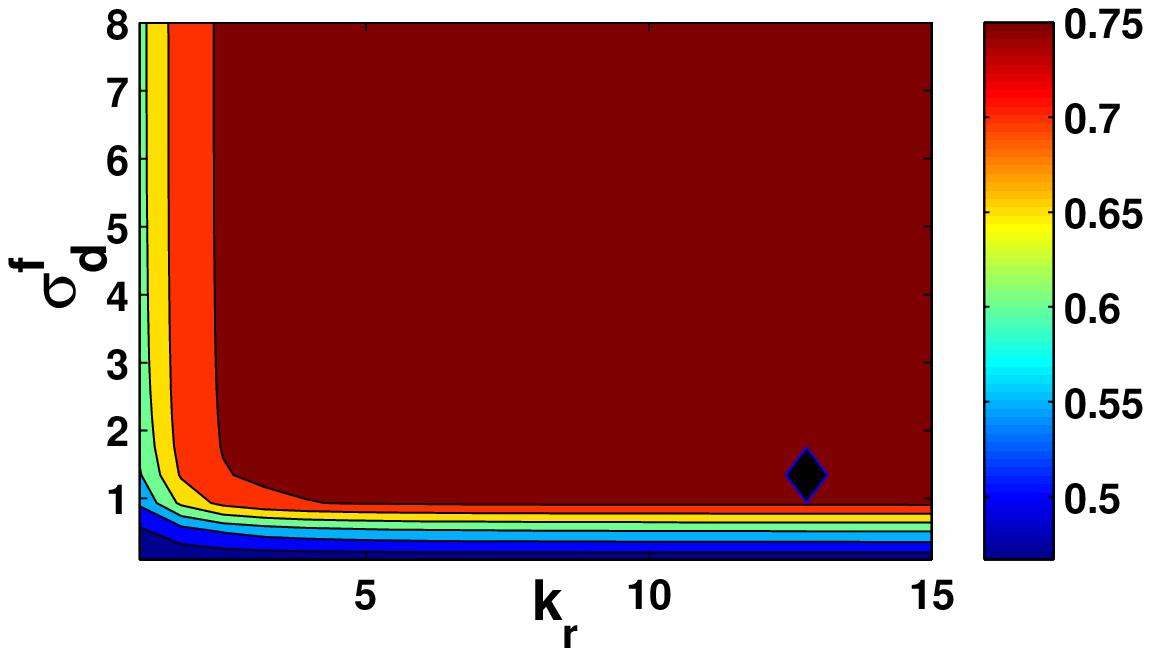}}
    \caption{\small{Tuning of BF parameters, $\sigma_d^f$ and $k_r$ to obtain maximum denoising performance. The mean PSNR measure at different values of $\sigma_d^f$ and $k_r$, when image is contaminated with AWGN of (a) $\sigma = 10$ and (b) $\sigma = 50$. Similarly, the SSIM measures is also obtained for image corrupted with AWGN of (c) $\sigma = 10$ and (d) $\sigma = 50$. The ``black diamond" symbol indicates peak value. The experiment was conducted on the reference images of TID2008 database \cite{ponomarenko2009tid2008}}}
    \label{Fig:Sec3.3:PSNR_SSIM_SigmaDR}
\end{figure*}

\section{Experimental Results \& Discussion}
\label{Sec:Results}
Most of the literatures are confined in using a particular type of image sets for the analysis of corresponding denoising algorithm/s \cite{luo2015adaptive}. Moreover, the image quality assessment (IQA) measures such as: Peak Signal to Noise Ratio (PSNR) and Structural Similarity Index Measure (SSIM) \cite{wang2004image} may not completely evaluate the performance of several image denoising techniques \cite{hore2010image,zeng2013perceptual}. Therefore, for completeness, in addition to the PSNR and SSIM indices, authors have considered the edge keeping index (EKI) to measure the edge strength (in terms of its magnitude) retained by the denoised image $\hat{z}$ compared to its true edges in $z$ as \cite{bhadauria2013medical}:

\begin{equation}\label{Eq:Sec4.3:EKI}
    EKI = \frac{\sum\limits_{\mathcal{P} = 1}^{M} \left[\Delta z(\mathcal{P}) - \Delta \mu_{z}\right] \left[\Delta \hat{z}(\mathcal{P}) - \Delta \mu_{\hat{z}}\right]}{\sqrt{\sum\limits_{\mathcal{P} = 1}^{M} \left[\Delta z(\mathcal{P}) - \Delta \mu_{z}\right]^2 \sum\limits_{\mathcal{P} = 1}^{M} \left[\Delta \hat{z}(\mathcal{P}) - \Delta \mu_{\hat{z}}\right]^2}}
\end{equation}

where, the reference image, $z$ and the denoised image, $\hat{z}$ are passed though high-pass filter (viz. Laplacian operator) to obtain $\Delta z$ and $\Delta \hat{z}$, respectively. Moreover, $\Delta \mu_{z}$ and $\Delta \mu_{\hat{z}}$ represent the mean of filtered output of original and restored image. A higher value of EKI -- in an interval $[0,1]$ --  indicates that the edge strength of original image is well preserved by the denoised image. In the following subsections, few parameters of the proposed algorithm is tuned empirically using above image quality assessment (IQA) indices for a wide variety of natural images, before accessing its performance, quantitatively.

\subsection{Parameter Optimization}
\label{SubSec:ParameterOpti}
The performance of proposed image denoising algorithm depends on the selection of both spatial and intensity domain parameters of JBF and BF in the coarser and the finest scale. The weights of the BF (and JBF) are multiplied and if one of the weights closes to zero, then no smoothing occurs. Thus choosing a narrower spatial kernel with larger range kernel produces limited smoothing. The dependencies among the kernels require combined tuning of $\alpha$ and $\sigma_d^c$ in the coarser scales for JBF (see Eq.\ref{Eq:Sec3:JBF_CoarserScales}) and $k_r$ and $\sigma_d^f$ in the finest scale for BF (see Eq.\ref{Eq:Sec3:BF_FinerScales}). We experimentally tuned these parameters in terms of maximum PSNR and SSIM measures at low ($\sigma = 10$) and high ($\sigma = 50$) noise levels, assuming the other parameters as given in Table.\ref{Table:SubSec_Optimal Parameter} are constant. The tuning of $\sigma_d^f$ and $k_r$ is shown in Fig.\ref{Fig:Sec3.3:PSNR_SSIM_SigmaDR} for TID2008 image database \cite{ponomarenko2009tid2008}. It may be observed that the optimal value of $\sigma_d^f$ is relatively independent to the noise variance $\sigma$ compared to $k_r$. On the other hand, the parameter $k_r$ changes significantly with $\sigma$. At lower noise strength $\sigma=10$ (see Fig.\ref{fig:PSNR_SigmaDR_TID10}), when $k_r$ is sufficiently large, increasing the spatial kernel parameter $\sigma_d$ over-smooth the coefficients. However, at higher noise strength $\sigma = 50$, when $k_r$ occupies larger value, the range kernel widens and flattens i.e. it becomes nearly constant over the intensity interval of the noisy coefficient. Therefore at higher $\sigma$ the change in denoising quality is insignificant with further change in $k_r$.

Literature suggests theoretically, it is difficult to select a single value of, $\sigma_r/\sigma$ (both in case of JBF and BF), which is optimal for all images and $\sigma_d$ value \cite{zhang2008multiresolution}. However, based on experimental observation, we fixed a single value for $\sigma_d$ for all noise levels. On the other hand, the range parameter\footnote{For JBF $\sigma_r^c$ is selected by varying $\alpha$ and for BF $\sigma_r^f = k_r \times \lambda_\gamma$ is selected by tuning the parameter $k_r.$}, $\sigma_r$ is selected according to the noise strength $\sigma$. Unlike, JBF and BF, the parameters $k$ and $k_1$ in Eq.\ref{Eq:Sec3:Threshold} and Eq.\ref{Eq:sec:GIF_SingleIm} are tuned independently to obtain maximum denoising quality. The final optimized values of these parameters are given in Table.\ref{Table:SubSec_Optimal Parameter}. In this article, the proposed image denoising technique is quantified and compared with several state-of-the-art techniques considering the default parameter values as given in Table.\ref{Table:SubSec_Optimal Parameter}.


\begin{table}[h]
\centering
\small
\caption{\small{Optimized Parameter Values for Proposed Image Denoising Framework.}}
\label{Table:SubSec_Optimal Parameter}
\begin{tabular}{|c|c|c|}
\hline
\textbf{Parameters}            & \textbf{Symbol}           & \textbf{Default Values}
\\ \hline
Thresholding Parameter                & $k$        & 2.0
\\ \hline
\multirow{2}{*}{\begin{tabular}[c]{@{}c@{}}JBF Kernels Shape \\ Control Parameters\end{tabular}} & $\alpha$         & 0.04
                                                                                                    \\ \cline{2-3}
                                                                                                 & $\sigma_{d}^{c}$ & 1.9  \\ \hline
\multirow{3}{*}{\begin{tabular}[c]{@{}c@{}}BF Kernels Shape \\ Control Parameters\end{tabular}}  & \multirow{2}{*}{$k_r$}            & 3.5 ($\sigma \le 40$)       \\ \cline{3-3}
                                                                                                 &                & 11 ($\sigma > 40$)       \\ \cline{2-3}
                                                                                                 & $\sigma_{d}^{f}$           & 1.27        \\ \hline
Regularization Parameter of GIF         & $k_1$    &  1.3 \\ \hline
\end{tabular}
\end{table}


\begin{figure*}[!ht]
    \centering
    \subfloat[{BF \cite{tomasi1998bilateral} \label{Fig:SubSec:BF_Lena}}]{\includegraphics[width=0.3\textwidth,natwidth=290,natheight=290]{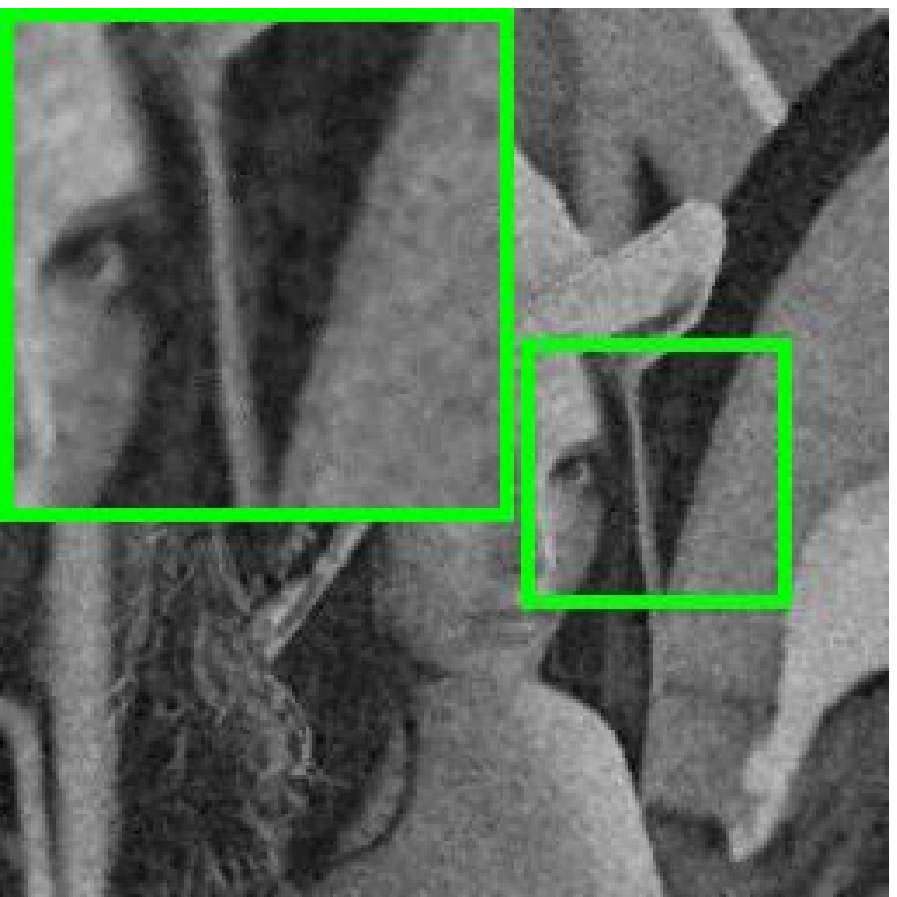}}
    \hspace{0.01cm}
    \subfloat[{CT \cite{starck2002curvelet} \label{Fig:SubSec:CTu_Lena}}]{\includegraphics[width=0.3\textwidth,natwidth=290,natheight=290]{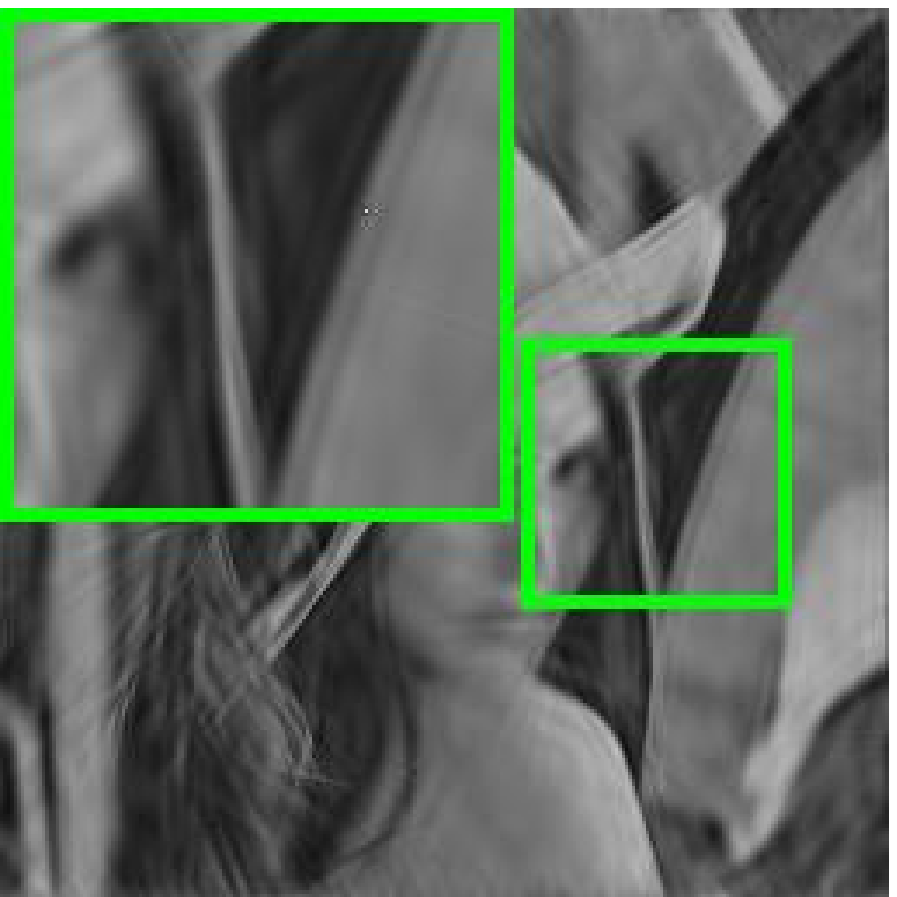}}
    \hspace{0.01cm}
    \subfloat[{KSVD \cite{elad2006image} \label{Fig:SubSec:KSVD_Lena}}]{\includegraphics[width=0.3\textwidth,natwidth=290,natheight=290]{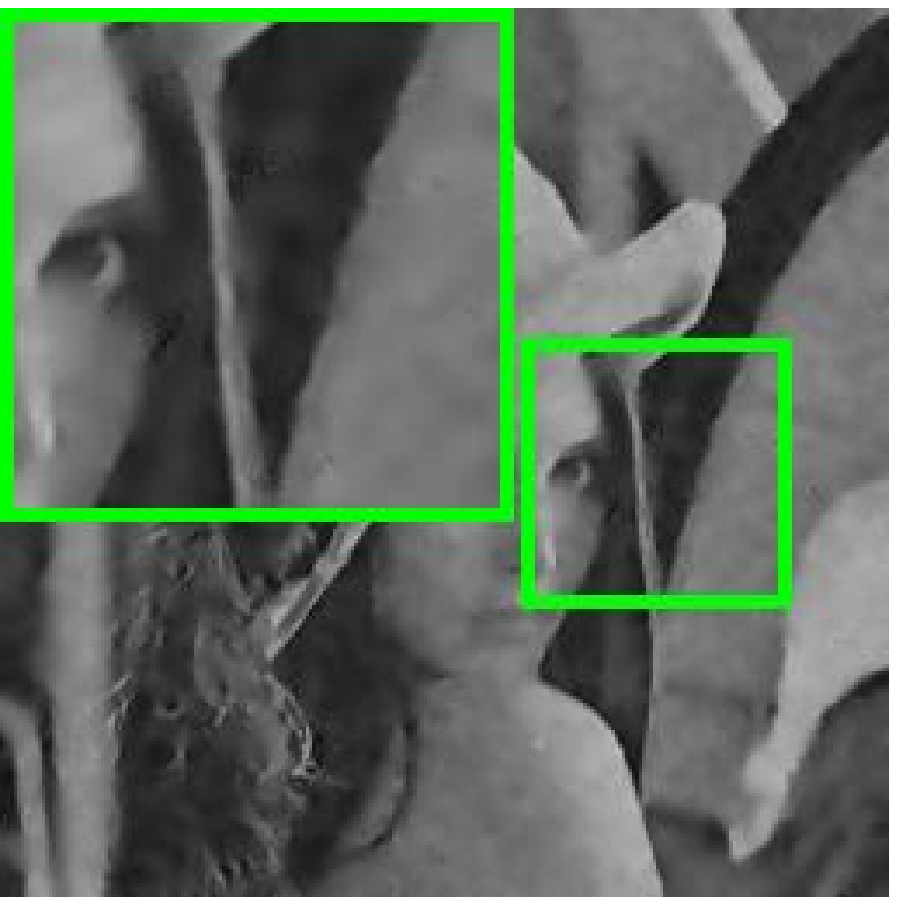}}\\
    \hspace{0.01cm}
    \subfloat[{NeighSURE \cite{dengwen2008image} \label{Fig:SubSec:NeighSURE_Lena}}]{\includegraphics[width=0.3\textwidth,natwidth=290,natheight=290]{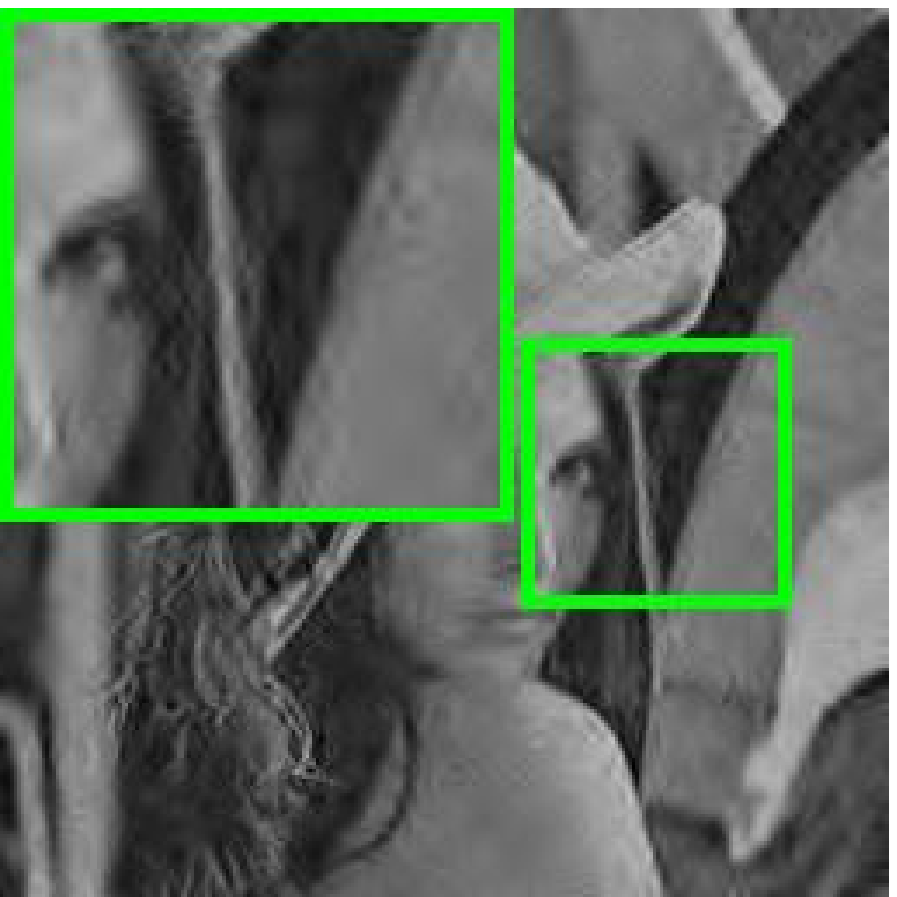}}
    \hspace{0.01cm}
    \subfloat[{MBF \cite{zhang2008multiresolution} \label{Fig:SubSec:MBF_Lena}}]{\includegraphics[width=0.3\textwidth,natwidth=290,natheight=290]{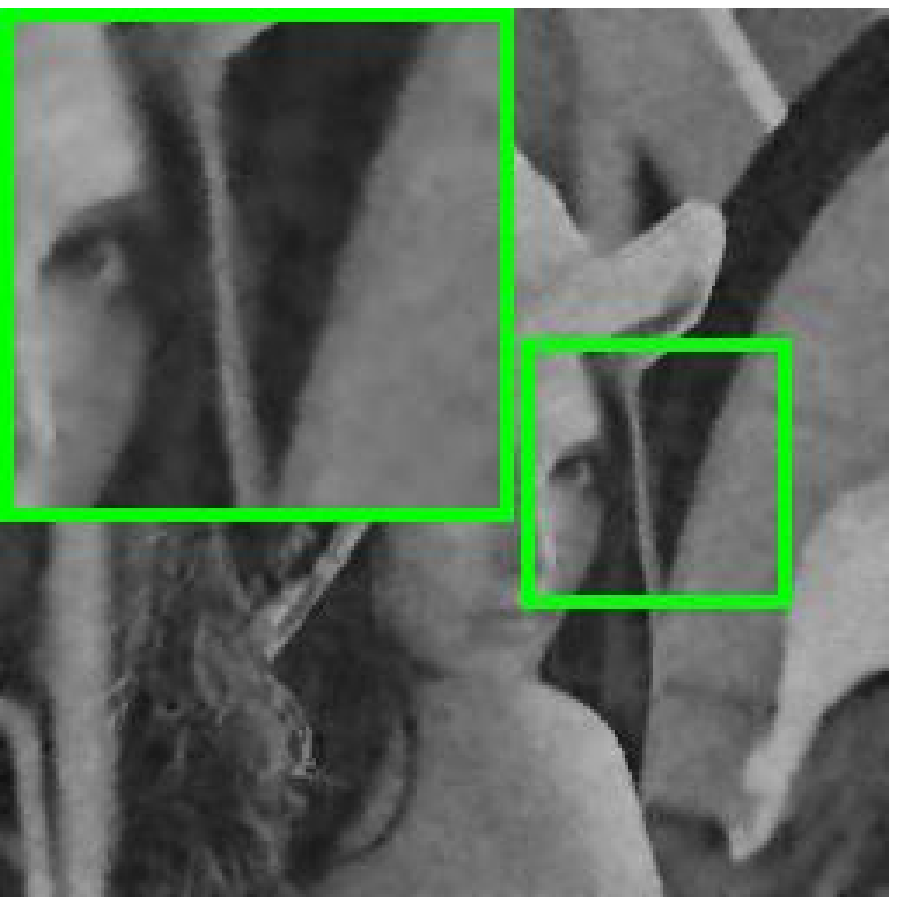}}
    \hspace{0.00cm}
    \subfloat[{NLM-SAP \cite{deledalle2012non} \label{Fig:SubSec:NLMSAP_Lena}}]{\includegraphics[width=0.3\textwidth,natwidth=290,natheight=290]{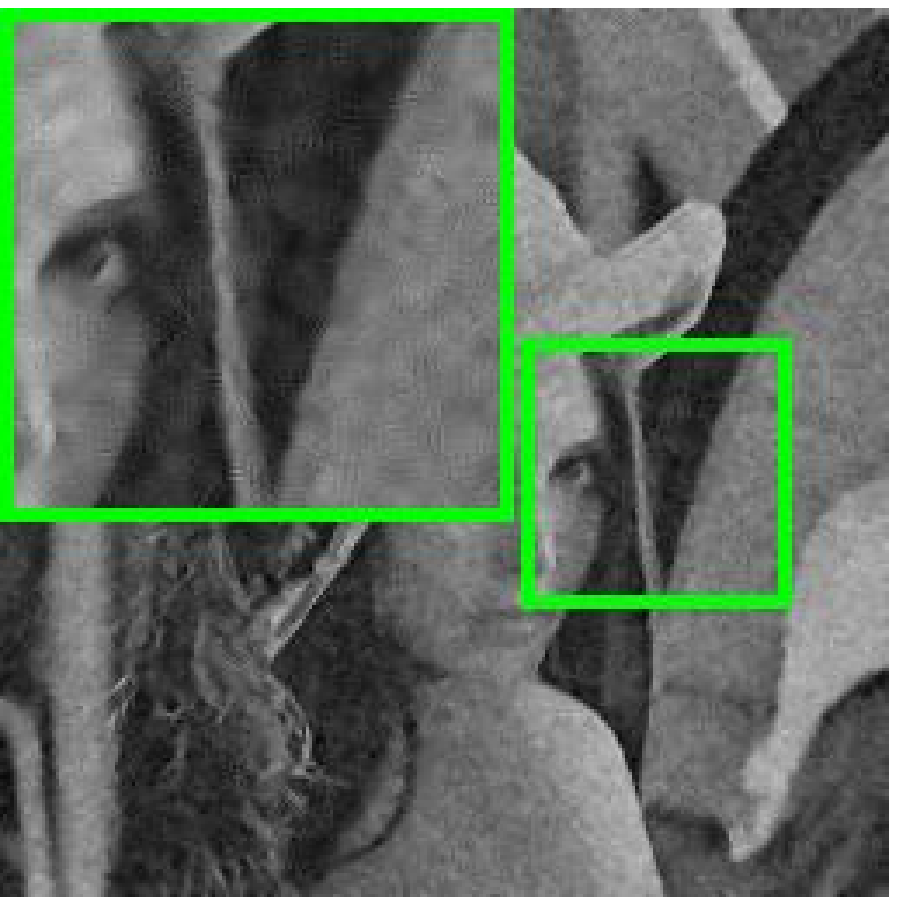}}\\
    \hspace{0.01cm}
    \subfloat[NLMNT \cite{kumar2013image}]{\includegraphics[width=0.3\textwidth,natwidth=290,natheight=290]{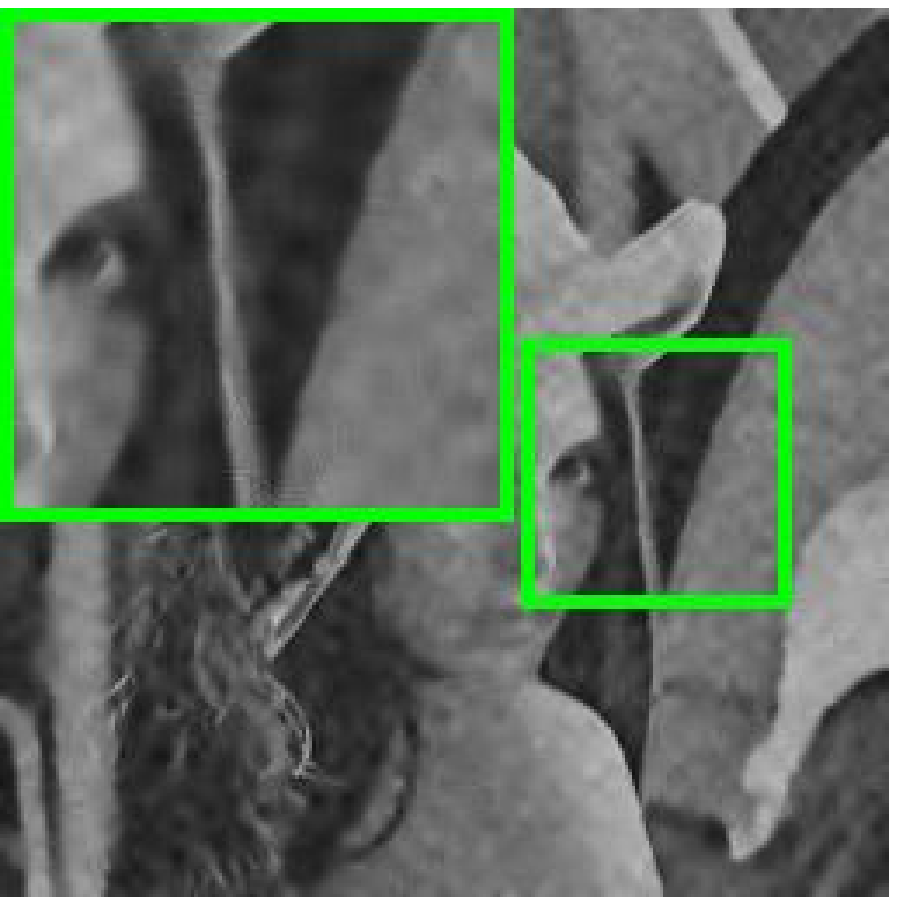}}
    \hspace{0.01cm}
    \subfloat[{BM3D \cite{dabov2007image} \label{Fig:SubSec:BM3D_Lena}}]{\includegraphics[width=0.3\textwidth,natwidth=290,natheight=290]{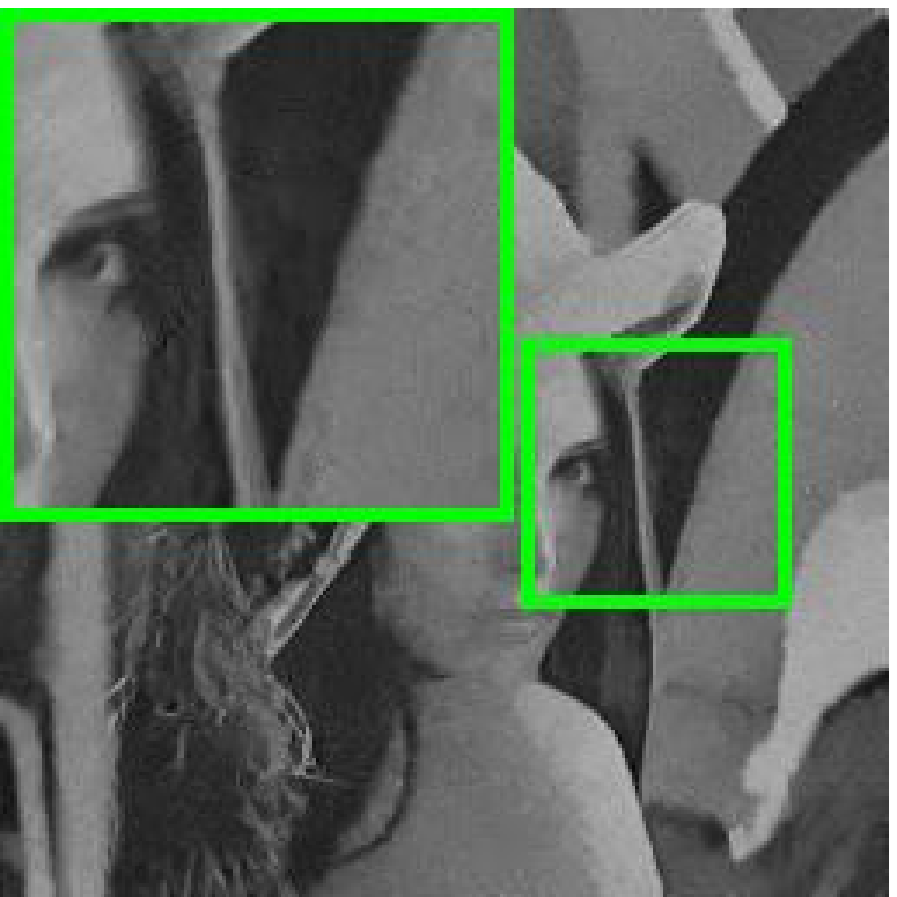}}
    \hspace{0.01cm}
    \subfloat[{Propose Method \label{Fig:SubSec:Proposed_Lena}}]{\includegraphics[width=0.3\textwidth,natwidth=290,natheight=290]{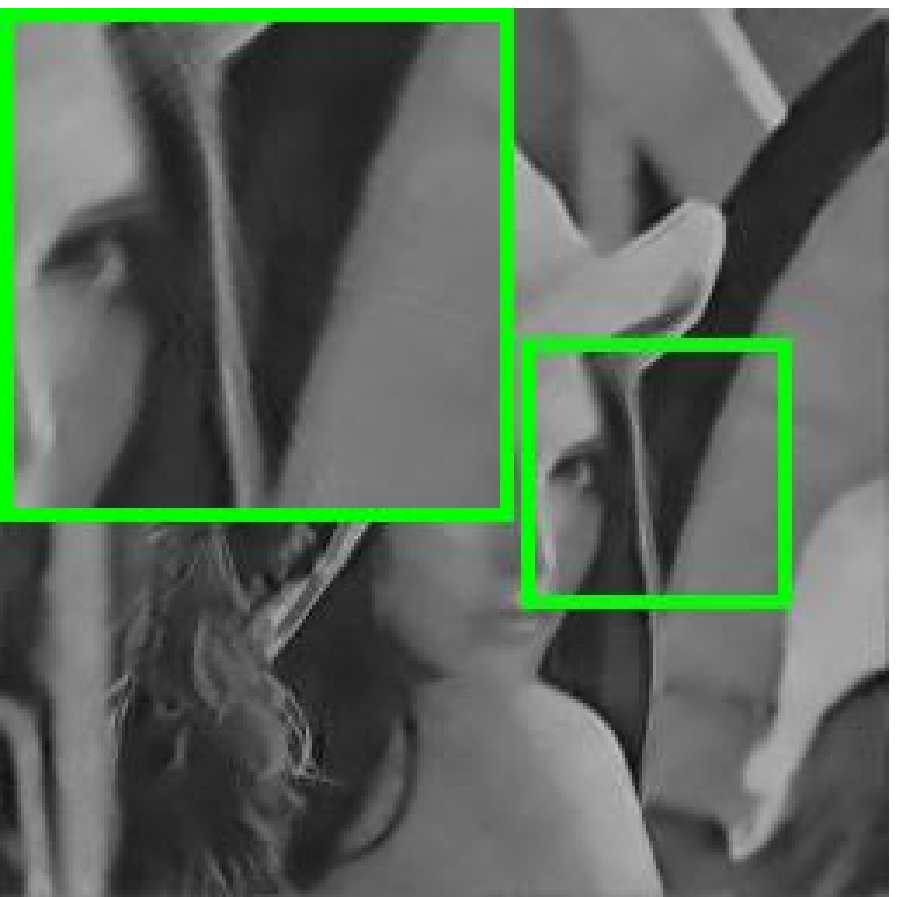}}
    \caption{\small{Denoised image obtained using different algorithms for Lena image corrupted with AWGN of $\sigma = 40$. The quantitative measures for respective denoised images are (a) PSNR = 27.899, SSIM = 0.745, EKI = 0.7923, (b) PSNR = 28.655, SSIM = 0.832, EKI = 0.889 (c) PSNR = 28.827, SSIM = 0.853, EKI = 898 (d) PSNR = 29.463, SSIM = 0.837, EKI = 0.900 (e) PSNR = 29.233, SSIM = 0.832, EKI = 0.891 (f) PSNR = 29.225, SSIM = 0.842, EKI = 0.886 (g) PSNR = 29.078, SSIM = 0.807, EKI = 881 (h) PSNR = 29.632, SSIM = 0.854, EKI = 0.846 and (i) PSNR = 30.001, SSIM = 0.862, EKI = 0.9101. A patch of each image is magnified for visual assessment.}}
    \label{Fig:subsecPerfEval:GrayImageDenoise}
\end{figure*}

\subsection{Denoising Performance Evaluation}
\label{SubSec:Denoise_Perform}
The efficacy of proposed algorithm is investigated for both artificial and natural images corrupted with additive white Gaussian noise (AWGN). Initially,the efficacy of proposed method -- in terms of statistical, structural and edge keeping index -- is compared with several state-of-the-art denoising techniques for Lena image corrupted with simulated additive noise of $\sigma = 40$. Fig.\ref{Fig:subsecPerfEval:GrayImageDenoise} illustrates the respective quantitative measures for the following methods: BF \cite{tomasi1998bilateral}, Curvelet Thresholding (CT) \cite{starck2002curvelet}, K-SVD \cite{elad2006image}, DWT-NeighSURE \cite{dengwen2008image}, MBF \cite{zhang2008multiresolution}, NLM-SAP \cite{deledalle2012non}, NLMNT \cite{kumar2013image}, BM3D \cite{dabov2007image}. A small patch of each denoised image is magnified (up to $300\%$) to indicate the performance of proposed technique in suppressing noise, while producing sharp edges with minimum loss of fine details even at higher noise strength ($\sigma = 40$).

The performance of our method was also investigated on a standard database with images that represent a wide variety of natural scenes. Here, the TID2008 database \cite{ponomarenko2009tid2008} -- that consists of $24$ natural images of different textural characteristic, homogeneous regions and edges -- is considered to evaluate the performance of several image denoising algorithms. Initially, all the reference images are converted to grayscale and then resized to $512 \times 512$ for uniformity in comparison. Table.\ref{Table:Sec5:Performance_Measure TID} yields the mean PSNR, SSIM and EKI measures for the reference images contaminated with AWGN of $\sigma = [10, 20, 30, 40, 50, 60, 70, 75]$. The following few facts may be observed from the comparative analysis:


\begin{table*}[!ht]
\centering
\caption{\small{Mean PSNR (dB), SSIM and EKI Measure between original and restored images for different denoising techniques on TID$2008$ Database \cite{ponomarenko2009tid2008}.}}
\label{Table:Sec5:Performance_Measure TID}
\begin{tabular}{|c|cccccccc|}
\hline
\multicolumn{1}{|c|}{Noise Level}   & \multicolumn{1}{c|}{$\sigma = 10$} & \multicolumn{1}{c|}{$\sigma= 20$} & \multicolumn{1}{c |}{$\sigma= 30$} & \multicolumn{1}{c|}{$\sigma = 40$} & \multicolumn{1}{c|}{$\sigma = 50$} & \multicolumn{1}{c |}{$\sigma = 60$} & \multicolumn{1}{c|}{$\sigma = 70$} & \multicolumn{1}{c|}{$\sigma = 75$} \\ \hline \hline
\multicolumn{9}{|r|}{\textbf{Peak Signal to Noise Ratio (PSNR) in dB}}                                                                                                                                                                                                                                                                             \\ \hline
\multicolumn{1}{|c|}{BF \cite{tomasi1998bilateral}} & 32.333                & 28.618                    & 26.708                            & 25.479                  & 24.577                             & 23.852                             & 23.238                             & 22.961                             \\
CT \cite{starck2002curvelet}                       & 31.028                             & 28.045                            & 26.555                            & 25.557                             & 24.823                             & 24.238                             & 23.754                             & 23.544                             \\
KSVD \cite{elad2006image}           & \textbf{34.593}                             & \textbf{30.892}           & 28.844                            & \textbf{27.405}                             & 26.305                             & 25.434                             & 24.719                 & 24.409               \\
NeighSURE \cite{dengwen2008image}               & 33.535                             & 29.740                            & 27.759                            & 26.474                             & 25.544                             & 24.828                             & 24.255                  & 24.001                       \\
MBF \cite{zhang2008multiresolution}               & 32.669                             & 29.413                            & 27.414                            & 26.245                             & 25.403                             & 24.751                             & 24.212                             & 23.972        \\
NLM-SAP \cite{deledalle2012non}               & 33.750                             & 30.190                            & 28.145                            & 26.688                             & 25.592                             & 24.744                             & 24.064                             & 23.768         \\
NLMNT \cite{kumar2013image}                   & 32.821                             & 29.822                            & 27.896                            & 26.379                             & 25.106                             & 24.015                             & 23.058                    & 22.621                \\
BM3D \cite{dabov2007image}                  & 34.537                             & 30.884                            & \textbf{28.863}                            & 27.192                             & \textbf{26.376}               & 25.530                & 24.810                             & 24.488                             \\
Proposed                             & 33.456                             & 30.186                            & 28.499                            & 27.049                             & 26.179                             & \textbf{25.701}                             & \textbf{25.073}                         &  \textbf{24.526} \\ \hline
\hline
\multicolumn{9}{|r|}{\textbf{Structural Similarity Measure \cite{wang2004image}}}                                                                                                                                                                                                                                                                              \\ \hline
BF \cite{tomasi1998bilateral}                      & 0.9264                             & 0.8550                            & 0.7859                            & 0.7274                             & 0.6768                             & 0.6319                             & 0.5914                             & 0.5725                             \\
CT \cite{starck2002curvelet}               & 0.9323                             & 0.8568                            & 0.7965                            & 0.7494                             & 0.7121                             & 0.6811                             & 0.6550                             & 0.6436                             \\
KSVD \cite{elad2006image}                   & \textbf{0.9608}               & 0.9053                            & 0.8519                            & 0.8032                             & 0.7589                             & 0.7188                             & 0.6827                             & 0.6659                             \\
NeighSURE \cite{dengwen2008image}       & 0.9544         & 0.8947              & 0.8402      & 0.7931       & 0.7521                             & 0.7161                             & 0.6859                             & 0.6713                             \\
MBF \cite{zhang2008multiresolution}                  & 0.9314                          & 0.8673                          & 0.8160                            & 0.7722                             & 0.7332                & 0.6975                   & 0.6642                  & 0.6483          \\
NLM-SAP \cite{deledalle2012non}      & 0.9504                         & 0.8806                            & 0.8196         &  0.7666   & 0.7200                             & 0.6788                             & 0.6420                             & 0.6250                             \\
NLMNT \cite{kumar2013image}      & 0.9532                             & 0.8955                            & 0.8336                            & 0.7740                             & 0.7187                             & 0.6688                             & 0.6240                             & 0.6035                             \\
BM3D \cite{dabov2007image}                  & 0.9605                             & \textbf{0.9096}                            & \textbf{0.8527}          & 0.8157                             & 0.7762                          & 0.7404                        & 0.7072                         & 0.6914                       \\
Proposed                              & 0.9547                             & 0.9122                            & 0.8712      & \textbf{0.8201}   & \textbf{0.7893}                & \textbf{0.7614}                      & \textbf{0.7238}               & \textbf{0.7014} \\ \hline
\hline
\multicolumn{9}{|r|}{\textbf{Edge Keeping Index \cite{bhadauria2013medical}}}                                                                                                                                                                                                                                                                                         \\ \hline
BF \cite{tomasi1998bilateral}           & 0.9395                             & 0.8954                            & 0.8644                            & 0.8416                             & 0.8237                             & 0.8086                             & 0.7952                             & 0.7889                             \\
CT \cite{starck2002curvelet}         & 0.9136                             & 0.8776                            & 0.8570                            & 0.8395                             & 0.8240                             & 0.8096                             & 0.7959                             & 0.7894                             \\
KSVD \cite{elad2006image}         & \textbf{0.9649}                & \textbf{0.9346}               & \textbf{0.9082}             & \textbf{0.8834}                             & 0.8598                             & 0.8377                             & 0.8170                             & 0.8073                             \\
NeighSURE \cite{dengwen2008image}     & 0.9534                             & 0.9141                            & 0.8848                            & 0.8624                             & 0.8442                             & 0.8295                             & 0.8174                             & 0.8116                             \\
MBF \cite{zhang2008multiresolution}          & 0.9474                             & 0.9077                            & 0.8770                            & 0.8521                             & 0.8312                   & 0.8129                             & 0.7991                             & 0.7912                             \\
NLM-SAP \cite{deledalle2012non}          & 0.9473                             & 0.9164                            & 0.8906                            & 0.8637                             & 0.8364                             & 0.8101                             & 0.7854                             & 0.7737                             \\
NLMNT \cite{kumar2013image}            & 0.9305                             & 0.8907                            & 0.8682                            & 0.8444                             & 0.8160                             & 0.7866                             & 0.7601                             & 0.7491                             \\
BM3D \cite{dabov2007image}         & 0.9620                             & 0.9272                            & 0.8913                            & 0.8386                             & 0.8346                             & 0.8012                             & 0.7672                             & 0.7501                             \\
Proposed                                & 0.9514                             & 0.9216                            & 0.8903                            & 0.8790                             & \textbf{0.8654}                 & \textbf{0.8524}              & \textbf{0.8328}                  & \textbf{0.8207} \\ \hline
\end{tabular}
\end{table*}

\begin{enumerate}
\item The estimation of signal below threshold using JBF aides in attenuating the sudden jumps in Curvelet magnitude and also preservers the essential phase information. The multiscale JBF in coarser scales and BF in the finest scale improves the performance of proposed technique compared to the individual methods of BF \cite{tomasi1998bilateral} and Curvelet Threholding (CT) \cite{starck2002curvelet}.
\item The approach of GIF as a post-processing `Edge-Aware' filter for the suppression of ringing artifacts and preservation of small image details aides in the enhancement of image quality. The improvement in EKI measures especially at higher noise levels indicate the importance of phase preservation\footnote{At higher noise strength $\sigma$, phase is less sensitive to noise as shown in Fig.\ref{PhaseSensitivityPlot_AWGN}.} using JBF and the localization of ringing artifacts due to GIF.
\item The multiscale hybrid image denoising technique excelled in performance at higher noise strengths compared to several state-of-the-art image denoising techniques \cite{elad2006image,dabov2007image,zhang2008multiresolution}. However its performance is comparable at lower noise levels.
\end{enumerate}

The intra-scale dependencies among the coefficients and the sparse recovery of any curve singularity during reconstruction motivated the authors to use Curvelet transform for image restoration. The multiscale (Joint Bilateral) filtering in Curvelet domain exploits the high co-dependency among the coefficients, while estimating signal residuals from noise subspace. The phases of the estimated magnitudes retain the essential location information of the image features. In contrast to multiscale filtering, spatial domain implementations fail to retain few image details in the low-contrast regions. It is also observed that the BF suffers from gradient reversal artifact near edges \cite{he2013guided}. Similarly, the NLM filters search for similar patches in a larger neighboring window, which is again local compared to the whole image \cite{buades2011non} and the similarity among patches further decreases with higher noise strength. Thus the performance of NLM-SAP \cite{deledalle2012non} and NLMNT \cite{kumar2013image} degrades significantly at higher values of $\sigma$. On the other hand, one may notice from Fig.\ref{Fig:subsecPerfEval:GrayImageDenoise} that the decimated wavelet transform -- in DWT-NeighSURE \cite{dengwen2008image} and MBF \cite{zhang2008multiresolution} -- yields distortions of the boundaries and suffers substantial loss of important image detail. A similar observation may also be noticed for Curvelet thresholding \cite{starck2002curvelet}. In addition, by setting higher threshold in Curvelet domain to avoid few of these distortion in the finest scale would cause even more of the inherent structure to be missed. We retained these structures using BF. The GIF, as illustrated in Fig.\ref{Fig:subsecAlgoGray:GIF} localizes the distortions of the boundaries, while keeping the edges, textures and small details of the latent image. Though the block-matching 3D (BM3D) collaborative filter improves the denoising quality significantly, it introduces few visible artifacts in the homogenous regions\footnote{The lower EKI measures validate the presence of low-frequency noise in the homogenous region.} as shown in Fig.\ref{Fig:SubSec:BM3D_Lena}. Further, in the absence of similar patches the performance of BM3D deteriorates at higher noise strengths. However, the competitiveness of proposed algorithm is comparable with several state-of-the-art techniques with maximum preservation of image details even at higher noise strength.


\section{Conclusion}
\label{Sec:Conclusion}
In this article, we proposed a multiscale filtering that improves the performance of classical Curvelet thresholding method for image denoising. The signal estimation in noise subspace using JBF and retention of phase seems to offer advantages in coarser scales, particularly in avoiding visible artifacts. The improvement in EKI measure justifies the importance of phase preservation during image restoration. Our experiments also reveal that Curvelet thresholding in the finest scale suffers substantial loss of important image detail. Therefore, removal of coefficients with thresholds introduces granular artifacts in the denoised image. The proposed BF ensures the elimination of granular artifacts with well-connected edges in the restored image. The reconstructed image is further processed by GIF to reduce the ringing artifacts, which is mostly ignored by many transform domain techniques. We empirically investigated the performance of proposed method under various noise strength to obtain optimum parameter values for JBF and BF in coarser and the finest scales, respectively. The competitiveness of proposed method is examined on TID2008 image database to emphasize its efficacy in diverse fields of applications. The experimental results illustrated the consistency of proposed algorithm compared to several state-of-the-art image denoising techniques.

\section*{Acknowledgement}
The authors would like to express their gratitude towards Prof. Prasanna K. Sahu, Department of Electrical Engineering, National Institute of Technology, Rourkela-08, for his gracious encouragement and support throughout this work.



%

{\small
\bibliographystyle{IEEEtran}
\bibliography{Ref_CT_JBF}}

\end{document}